\documentclass[11pt,a4paper]{article}

\usepackage{graphicx}
\usepackage{epsfig}

\usepackage{hyperref}
\usepackage{color}
\usepackage{pifont}
\usepackage{ mathrsfs }

\usepackage{graphicx,epsf}
\usepackage{bm}
\usepackage{amsmath}

\usepackage{listings}

\definecolor{red}{rgb}{0.8, 0., 0.}
\definecolor{darkgreen}{rgb}{0.,0.5, 0.}

\newcommand{\aeq}{\begin{equation}}
\newcommand{\eeq}{\end{equation}}
\newcommand{\aeqn}{\begin{eqnarray}}
\newcommand{\eeqn}{\end{eqnarray}}
\newcommand{\aeqs}{\begin{equation*}}
\newcommand{\eeqs}{\end{equation*}}
\newcommand{\aeqns}{\begin{eqnarray*}}
\newcommand{\eeqns}{\end{eqnarray*}}
\newcommand{\vecb}[1]{{\bf #1}}

\newcommand{\fpar}[2]{\frac{\partial{#1}}{\partial{#2}}}

\def \d {\mathrm{d}}
\begin{document}

\begin{center}
{\LARGE{\textbf{Linear gyrokinetic particle-in-cell simulations of Alfv\'en instabilities in tokamaks.}}}\\
\vspace{0.2 cm}
{\normalsize {\underline{A. Biancalani}${}^{1}$, A. Bottino${}^{1}$, S. Briguglio${}^{2}$, A. K\"onies${}^{3}$, Ph. Lauber${}^{1}$, A. Mishchenko${}^{3}$, E.Poli${}^{1}$, B. D. Scott${}^{1}$, F. Zonca${}^{2,4}$.}}\\
\vspace{0.2 cm}
\small{${}^1$ Max-Planck-Institut f\"ur Plasmaphysik, 85748 Garching, Germany\\
${}^2$ ENEA C. R. Frascati, Via E. Fermi 45, CP 65-00044 Frascati, Italy\\
${}^3$ Max-Planck-Institut f\"ur Plasmaphysik, 17491 Greifswald, Germany\\
${}^4$ Institute for Fusion Theory and Simulation, Zhejiang Univ., Hangzhou, People's Rep. of China\\
\vspace{0.6 cm}
\url{www2.ipp.mpg.de/~biancala}
}
\end{center}

\begin{abstract}
The linear dynamics of Alfv\'en modes in tokamaks is investigated here by means of the global gyrokinetic particle-in-cell code ORB5, within the NEMORB project. The model equations are shown and the local shear Alfv\'en wave dispersion relation is derived, recovering the continuous spectrum in the incompressible ideal MHD limit.
A verification and benchmark analysis is performed for continuum modes in a cylinder and for toroidicity-induced Alfv\'en Eigenmodes.
Modes in a reversed-shear equilibrium are also investigated, and the dependence of the spatial structure in the poloidal plane on the equilibrium parameters is described. In particular, a phase-shift in the poloidal angle is found to be present for modes whose frequency touches the continuum, whereas a radial symmetry is found to be characteristic of modes in the continuum gap.
\end{abstract}

\section{Introduction.}
\label{sec:intro}

\vspace{-0.2cm}

Plasma heating is essential for reaching appropriate fusion temperatures in tokamak plasmas, but as a side effect global modes can become unstable by converting particle kinetic energy into collective kinetic energy. The energetic particle (EP) population produced in the process of heating together with the alpha particles produced in fusion reactions are important actors in this chain,  driving plasma oscillations unstable via resonant wave-particle interactions.
On the other hand, the plasma instabilities such as Alfv\'en Eigenmodes (AE) can redistribute the EP population making the plasma heating less effective, and leading to additional loads on the walls~\cite{Cheng85,Chen07,Mishchenko09,Lauber13}.
Typically, tokamak plasmas are turbulent plasmas. Thereby, turbulence adds one more level of difficulty to the EP-redistribution problem, where wave-wave nonlinear interaction of turbulent and zonal modes with AE competes with wave-particle nonlinear saturation mechanisms of AE. For these reasons, it is important to have a proper selfconsistent theoretical framework to understand the AEs' instability threshold in present tokamaks and predict it in future fusion reactors.

NEMORB~\cite{Bottino11,Bottino15JPP,Biancalani16CPC} is the project of development of an electromagnetic, multi-species version of the nonlinear gyrokinetic particle-in-cell (PIC) code ORB5~\cite{Jolliet07}. The Lagrangian formulation that is used, is based on the gyrokinetic GK Vlasov-Maxwell equations of Sugama, Brizard and Hahm~\cite{Sugama00,Brizard07}.
Due to the method of derivation of the GK Vlasov-Maxwell equations from a discretized Lagrangian, the symmetry properties of the starting Lagrangian are passed to the Vlasov-Maxwell equations, and the conservation theorem for the energy is automatically satisfied~\cite{Bottino15JPP}.
As a consequence, this model can be adopted in principle for rigorous nonlinear electromagnetic simulations of global instabilities in the presence of EP and turbulence, where all nonlinearities are treated on the same footing in a self-consistent way. Furthermore, a PIC formulation offers a fine discretization in v-space ``for free'', which is crucial for studying the wave-particle interaction in the narrow layers around the resonances in phase-space.

When trying to solve the Vlasov-Maxwell set of equations in a $p_\|$ formulation with a $\delta f$ PIC method, one faces a numerical problem called ``cancellation problem''~\cite{Mishchenko04,Hatzky07}. This arises in particular in the numerical resolution of the Amp\`ere's equation, i.e. in the equation for the vector potential. One term of this equation, namely the current integral, has to be calculated with a discretization in terms of macro-particles, i.e. markers, whereas the other terms are calculated directly as analytic integrals in phase-space (see Sec.~\ref{sec:model}).
Due to fact that the statistical error affects only the term discretized with markers,  the balance between these terms is not satisfied, and the result is a numerical error which can be orders of magnitude higher than the desired signal. 

A brute-force solution to the cancellation problem is the drastic increase in the number of markers, which in turn would make electromagnetic PIC simulations unpractical. A smart solution has been proposed as a split of the adiabatic and nonadiabatic parts of the electron distribution function, where only the physically relevant one, i.e. the nonadiabatic part, is discretized with markers, whereas the adiabatic part (which is dominant in absolute value) can be calculated directly by means of an adjustable control variate~\cite{Mishchenko04,Hatzky07}.
This scheme has been found to greatly mitigate the cancellation problem, making electromagnetic PIC simulations feasible, with a reasonable number of markers. Recently, the control-variate scheme has been implemented also in NEMORB~\cite{Bottino11,Biancalani16CPC}, making the investigation of the dynamics of shear Alfv\'en waves (SAW) possible, for linear and nonlinear simulations.
Furthermore, a split of the vector potential into symplectic and Hamiltonian parts can be performed, and only the Hamiltonian part has to be calculated selfconsistently with the Vlasov-Maxwell system of equations, whereas the symplectic part can be evaluated with alternative methods (for example by imposing the ideal MHD Ohm's law, which is valid for incompressible SAW to the leading order)~\cite{Mishchenko14}.
This new ``pullback'' scheme, which further helps in strongly mitigating the cancellation problem, is also considered as one of the next numerical improvements to be done in NEMORB.

NEMORB has been recently verified and benchmarked in electrostatic mode for linear dynamics of global instabilities driven by EP~\cite{BiancalaniNuFu14,ZarzosoNuFu14}, but a detailed verification and investigation of the linear dynamics of Alfv\'en instabilities has not yet been done with this code. This is a necessary step in the direction of performing a trustable study of the richer nonlinear dynamics of Alfv\'en modes in the presence of turbulence and zonal flows. In this paper, we firstly describe a verification and benchmark effort of NEMORB on Alfv\'en instabilities driven by EP.
Secondly, we investigate the effect of the continuum on the radial structure of Alfv\'en modes, finding that the radial symmetry is broken, when the mode frequency lies outside the continuum gaps.

The structure of the paper is the following: in Section~\ref{sec:model} we describe the model equations of NEMORB and derive the local dispersion relation of SAW, which constitutes the continuous spectrum. The continuous spectrum is the frequency, varying continuously in space, where Alfv\'en instabilities are damped due to continuum damping~\cite{Alfven42,Alfven50,Grad69,Chen74}. 
Generally speaking, the continuous spectrum does not exist in pure kinetic theory. But there exists a nearly dense spectrum of modes that behaves as a continuum when one considers their cumulative impact on a given oscillation, with assigned frequency and wave-vector~\cite{Zonca14b}. In this paper, we focus on the derivation of the continuum in the incompressible ideal MHD limit, by neglecting the effect of the EP. 
A verification of NEMORB for the frequency of the continuous spectrum is performed in the limit of very small inverse aspect ratio (i.e. in the cylindrical limit), and described in Section~\ref{sec:verification-cylinder}, in the absence of EP.
Toroidicity-induced AE (TAE) are investigated in the presence of EP in Sec.~\ref{sec:TAE}, and results compared with those obtained with the hybrid MHD-gyrokinetic code HMGC~\cite{Briguglio95} and published in Ref.~\cite{Koenies12}.
Sec.~\ref{sec:EPM} is devoted to a study of how EP drive continuum modes unstable, and in particular of the dependence of their spatial structure on the continuum properties.
Finally, Sec.~\ref{sec:conclusions} is devoted to a summary of the conclusions and an outline of the next steps.

\section{The model and the local dispersion relation of SAW.}
\label{sec:model}

\subsection{Model equations.}

In this section, we describe the general model equations of NEMORB, and we derive the local dispersion relation for the SAW continuous spectrum, in the limit of a cold plasma and neglecting the parallel equilibrium current $J_{\|0}$.
 
The gyrokinetic equation in its general form is described by the Liouville theorem, i.e. the property of incompressibility of the distribution function in phase-space, in the absence of collisions:
\begin{equation}
\frac{\partial f }{\partial t} + \frac{1}{B_\|^*}  \frac{\partial }{\partial \vecb{Z}} \cdot \big( B_\|^* \dot{\vecb{Z}} f \big) = 0 \label{eq:gk}
\end{equation}
The phase-space coordinates are $\vecb{Z}=(\vecb{R},p_\|,\mu)$, i.e. respectively the gyrocenter position, canonical parallel momentum $p_\| = m U + (e/c) J_0 A_\|$ and magnetic momentum $\mu = m v_\perp^2 / (2B)$. The Jacobian is given by the parallel component of $\vecb{B}^*= \vecb{B} + (c/e) p_\| \vecb{\nabla}\times \vecb{b}$, where $\vecb{B}$ and $\vecb{b}$ are the equilibrium magnetic field and magnetic unitary vector.

The properties of the system are described by the gyrokinetic Lagrangian (see Ref.~\cite{Bottino15JPP} and references therein):
\begin{eqnarray}
\mathscr{L} & = & \Sigma_{\rm{sp}}\int \d V \d W \Big[ \Big( \Big( \frac{e}{c}\vecb{A}+p_\|\vecb{b} \Big)      \cdot\dot{\vecb{R}}+\frac{m  c}{e}\mu \dot{\theta} \Big) f \nonumber \\
&& -\big( \mathscr{H}_0 + \mathscr{H}_1 \big) f -  \mathscr{H}_2 f_M \Big] - \int \d V \frac{|\nabla_\perp A_\parallel|^2}{8\pi} \label{eq:Lagrangian}
\end{eqnarray}
with the Hamiltonian divided into unperturbed, linear, and nonlinear part, $\mathscr{H} = \mathscr{H}_0 +\mathscr{H}_1 + \mathscr{H}_2 $, with:
\begin{eqnarray}
 \mathscr{H}_0 & = & \frac{p_\|^2}{2m} + \mu B \\
 \mathscr{H}_1 & = & e \Big(  J_0\Phi - \frac{p_\|}{mc} J_0 A_\parallel \Big)  \\
 \mathscr{H}_2 & = & \frac{e^2}{2mc^2} (J_0 A_\parallel)^2  - \frac{mc^2}{2B^2}|\nabla\phi|^2
\end{eqnarray}
Here $f$ and $f_M$ are the total and equilibrium (i.e. independent of time) distribution functions, the integrals are over the phase space volume, with $\d V$ being the real-space infinitesimal and $\d W = (2\pi/m^2) B_\|^* \d p_\| \d \mu$ the velocity-space infinitesimal.
 The time-dependent fields are the scalar potential $\phi$ and the parallel component of the vector potential $A_\|$.
Here $\vecb{A}$ is the equilibrium vector potential.  The summation is over all species present in the plasma, and the gyroaverage operator is labeled here by $J_0$ (with $J_0=1$ for electrons). The gyroaverage operator reduces to the Bessel function if we transform into Fourier space.
The Lagrangian given in Eq.~\ref{eq:Lagrangian} is a function of the trajectories and of the fields, and contains all information about the system we are interested in. In the following, the particle trajectories and field equations are derived from Eq.~\ref{eq:Lagrangian} (see also Ref.~\cite{Sugama00,Lee01,Miyato09,Scott10}).

The particle gyrocenter trajectories are derived by imposing the minimal action principle with respect to the phase-space coordinates, which yields~\cite{Bottino15JPP}\footnote{The sign of the second term at the R.H.S. appears wrong in the original reference, due to a typographical error.}:
\begin{eqnarray}
\dot{\vecb R}&=&\fpar{\big(\mathscr{H}_0+\mathscr{H}_1 \big)}{p_\|}\frac{\vecb{B^*}}{B^*_\parallel}+ \frac{c}{eB^*_\parallel}\vecb{b}\times\nabla \big(\mathscr{H}_0+\mathscr{H}_1 \big) \nonumber \\
\dot{p_\|}&=&-\frac{\vecb{B^*}}{B^*_\parallel}\cdot\nabla \big(\mathscr{H}_0+\mathscr{H}_1 \big) \nonumber
\end{eqnarray}
Now by noting that $\nabla\big(\mathscr{H}_0+\mathscr{H}_1 \big) = \mu \nabla B + e \nabla J_0 (\phi - A_\| p_\|/mc)$, we can write the trajectories in explicit form:
\begin{eqnarray}
\dot{\vecb  R}&=&\frac{1}{m}\left(p_\|-\frac{e}{c}J_0A_\parallel\right)\frac{\vecb{B^*}}{B^*_\parallel} + \frac{c}{e B^*_\parallel} \vecb{b}\times \left[\mu
  \nabla B + e \nabla J_0  \big(\phi -  \frac{p_\|}{mc} A_\| \big) \right] \label{eq:trajectories_a} \\
\dot{p_\|}&=&-\frac{\vecb{B^*}}{B^*_\parallel}\cdot\left[\mu \nabla B + e
  \nabla J_0 \big(\phi -  \frac{p_\|}{mc} A_\| \big) \right] \label{eq:trajectories_b}
\end{eqnarray}

The Poisson equation is derived by imposing the minimal action principle with respect to the scalar potential, which yields~\cite{Bottino15JPP}:
\begin{equation}
 - \vecb{\nabla} \cdot \frac{n_0 mc^2}{B^2} \nabla_\perp \phi=\Sigma_{\rm{sp}} \int \d W  e J_0 f  \label{eq:Poisson}
\end{equation}
with $n_0 m$ being here the total plasma mass density.

Similarly, the Amp\`ere equation is derived by imposing the minimal action principle with respect to the vector potential, which yields~\cite{Bottino15JPP}:
\begin{eqnarray}
\Sigma_{\rm{sp}} \int \d W \Big( \frac{ep_\|}{mc} f-\frac{e^2}{mc^2}A_\parallel f_{M}
 \Big)  +  \frac{1}{4\pi}\nabla_\perp ^2 A_\parallel =0 \label{eq:Ampere}
\end{eqnarray}
Here the form with $J_0=1$ is given for simplicity. For more complicated models, see Ref.~\cite{Bottino15JPP}.

As mentioned in Sec.~\ref{sec:intro}, the resolution of Eq.~\ref{eq:Ampere} presents a numerical problem, if one wants to solve the first term like it is written here, with a particle-in-cell technique.
This is because the total (equilibrium + perturbed) electron distribution function $f_e$ has to be integrated in the phase space by means of a marker discretization, whereas the term involving the equilibrium distribution function is solved by direct integration, yielding the equilibrium electron density.
Due to the fact that each of these two terms is much bigger in amplitude than their difference, and that the statistical error introduced by the marker discretization falls on the first term only, and not on the second one, then we have that the resulting balance (or ``cancellation'') is not satisfied in the numerical solution, which is dominated by the statistical noise~\cite{Mishchenko04,Hatzky07}.
A solution to this ``cancellation problem'' comes with a control-variate technique, which splits the perturbed distribution function $\delta f$ in an adiabatic part $\delta f^{ad} = - (J_0 \phi - p_\| J_0 A_\|/mc) e f_M / k_B T$ and a nonadiabatic part (i.e. the remaining part). With this technique, the integral to be performed with the marker discretization becomes in fact much smaller, and therefore the resulting numerical noise is greatly mitigated~\cite{Mishchenko04,Hatzky07}.
This control-variate technique has been recently implemented in NEMORB~\cite{Bottino11,Biancalani16CPC}, and allows us to perform the first numerical simulations of SAW.

Eqs.~\ref{eq:trajectories_a},~\ref{eq:trajectories_b},~\ref{eq:Poisson},~\ref{eq:Ampere} are the constitutive equations of the model. The results of NEMORB simulations, where these equations are solved numerically with particle-in-cell method, are described in Sec.~\ref{sec:verification-cylinder}, \ref{sec:TAE}, \ref{sec:EPM}.

\subsection{Local dispersion relation.}


In the following, we take Eqs.~\ref{eq:trajectories_a},~\ref{eq:trajectories_b},~\ref{eq:Poisson},~\ref{eq:Ampere} as starting point to derive analytically the vorticity equation (see also Ref.~\cite{Lee01,Miyato09} for analogous derivations).
We focus here on radially localized perturbations in the incompressible ideal MHD limit in a tokamak with large aspect-ratio, and  neglect the parallel equilibrium current. The shear-Alfv\'en wave continuous spectrum is derived as a result, in tokamak geometry.   

In order to derive the vorticity equation, we start by taking the time derivative of Poisson equation for $T=0$, and writing it in a continuity form~\cite{Miyato09}:
\begin{equation}
\frac{\partial w}{\partial t} + \nabla \cdot \vecb{J}_G = 0 \label{eq:vorticity_1}
\end{equation}
where the vorticity is defined by:
\begin{equation}
w =  - \vecb{\nabla} \cdot \frac{n_0 mc^2}{B^2} \nabla_\perp \phi
\end{equation}
and where the gyrocenter current is:
\begin{equation}
 \vecb{J}_G = \Sigma_{\rm{sp}} \int \d W  e f \dot{\vecb{R}}  \label{eq:gk-current}
\end{equation}
Here the gyrokinetic equation, Eq.~\ref{eq:gk}, has been used to give an explicit form to the time derivative of the distribution function (only the term with spatial derivatives survives inside the phase-space integral, due to the divergence theorem).

For the present derivation, we consider only the dominant term in the gyrocenter velocity. For a magnetized plasma, this is the one along the equilibrium magnetic field:
\begin{equation}
\dot{\vecb  R}\simeq \frac{1}{m}\left(p_\|-\frac{e}{c}J_0A_\parallel\right) \vecb{b} 
\end{equation}
The linearized divergence of the gyrocenter current becomes:
\begin{equation}
\nabla \cdot \vecb{J}_G \simeq \vecb{B}\cdot\nabla \Big[ \frac{1}{B}  \Sigma_{\rm{sp}} \int \d W \frac{e}{m} \Big( p_\| \delta f -  \frac{e A_\|}{c}f_M  \Big) \Big] = - \frac{c}{4\pi} \vecb{B}\cdot\nabla \Big( \frac{1}{B} \nabla_\perp ^2 A_\parallel \Big)
\end{equation}
where we have used Amp\`ere's equation, Eq.~\ref{eq:Ampere}, to evaluate the integral in phase-space, in the limit of a cold plasma ($J_0(k_\perp \rho_i)$ = 1).

%
%

We now consider a plasma where the parallel electric field is zero, $E_\| = -\vecb{b}\cdot \nabla \phi - (1/c) \partial_t A_\| = 0$, consistently with Ohm's law in the ideal MHD regime.
A Fourier transform is performed in time, so that $\partial_t \rightarrow -i \omega$, and the Ohm's law can be written as $i(\omega/c)  A_\| = \vecb{b}\cdot \nabla \phi$. A tokamak with circular concentric flux surfaces is considered in the following. After performing a time derivative to the whole equation, the vorticity equation for incompressible shear-Alfv\'en waves reads~\cite{Cheng85,Fu89}:
\begin{equation}\label{eq:vorticity_3}
\vecb{\nabla} \cdot \frac{ \omega^2 }{v_A^2} \nabla_\perp \phi + (\vecb{b}\cdot \nabla) \nabla_\perp^2 (\vecb{b}\cdot \nabla) \phi = 0
\end{equation}
where $v_A^2(\vecb{R}) = B^2/4\pi m n_0$ is the local Alfv\'en velocity. The singular solutions of this equation are the continuum modes, and reflect the fact that a global mode approaching the position of the continuum, is damped by continuum damping~\cite{Chen74}. The importance of knowing the exact topology of the continuous spectrum is clear, for the existence of global Alfv\'en instabilities has a strong dependence on whether their frequency touches or not the continuum.

In the following, we derive the SAW continuous spectrum formula in the tokamak geometry.
The coordinates used are the cylindrical ($R,\varphi,Z$) and toroidal
($r,\zeta,\theta$) coordinates linked by:
\begin{displaymath} 
R=R_0 + r \cos\theta
, \quad Z= r\sin\theta , \quad \varphi = -\frac{\zeta}{R}
\end{displaymath}
We consider a geometry with small inverse aspect ratio $\epsilon = a/R \ll 1$ for this derivation. The plasma nonuniformity is kept now only in the spatial dependence of the Alfv\'en velocity: $v_A(r,\theta) = v_A(r)/(1+\epsilon \cos\theta)$.
The modes are Fourier-decomposed in the poloidal and toroidal angles:
\begin{displaymath}
 \phi (r,\theta,\varphi,t) = \sum_m \phi_m(r) \exp{i(-m\theta + n\varphi
- \omega t)}
\end{displaymath}
Now we select one single toroidal mode, and the different components in m have:
\begin{displaymath}
k_{\parallel m}= \frac{1}{R_0}\Big(n - \frac{m}{q(r)} \Big)
\end{displaymath}
We write the vorticity equation, Eq.~\ref{eq:vorticity_3} as a matrix equation $M_{ij}(r,n,\omega) \phi_j = 0$, whose determinant is imposed to be zero. With the chosen decomposition in Fourier, it is clear that the plasma nonuniformity of the Alfv\'en velocity couples the $m$ and $m\pm 1$ modes. This reflects in the fact that the matrix has only diagonal terms and first off-diagonal terms nonzero, with value:
\begin{eqnarray}
M_{ii}(r,n,\omega)  & = &  \tilde\omega^2 - \tilde{k}^2_{\|m}\\
M_{ij}(r,n,\omega)  & = &  \hat\epsilon\; \tilde\omega^2   \; \; \;\;\;\;\; (j = i\pm1) \\
M_{ij}(r,n,\omega)  & = &  0   \; \; \;\;\;\;\;\;\;\;\; (|j-i| > 1)
\end{eqnarray}
where $\tilde\omega = q R \omega / v_A$, $\tilde{k}_{\|m} = qR k_{\|m}$, and $\hat\epsilon = 3r/2R$. When we focus on a region of the plasma in the proximity of the crossing of an $m$ and an $m+1$ continuum branches, the determinant reduces to a second-order algebraic equation, where the two solutions are:
\begin{equation}\label{eq:continuum-formula}
\tilde\omega^2_\pm =  \frac{\tilde{k}^2_{\|m}+\tilde{k}^2_{\|m+1}  \pm \sqrt{(\tilde{k}^2_{\|m}-\tilde{k}^2_{\|m+1})^2 + 4 \hat\epsilon^2 \tilde{k}^2_{\|m} \tilde{k}^2_{\|m+1} }   }{2 (1-\hat\epsilon^2)}
\end{equation}
This is the formula for the continuous spectrum of SAW in a tokamak geometry~\cite{Cheng85,Fu89}. In the cylindrical limit, $\epsilon \rightarrow 0$, Eq.~\ref{eq:continuum-formula} reduces to the local SAW dispersion relation $\omega^2 = v_A^2 k^2_{\|m}$. For a small but finite value of $\hat\epsilon$, a gap is created in the continuum at the position of the crossing of the two continuum branches with $m$ and $m+1$ (where $k_{\|m}=-k_{\|m+1}$). In this gap, global modes named ``toroidicity induced Alfv\'en Eigenmodes'' (TAE) can exist, essentially not damped by continuum damping.

In the next sections, the model equations of NEMORB, Eqs.~\ref{eq:trajectories_a},~\ref{eq:trajectories_b},~\ref{eq:Poisson},~\ref{eq:Ampere}, are solved numerically with NEMORB, i.e. with a particle-in-cell technique, and the results are compared with the continuum formula, Eq.~\ref{eq:continuum-formula}.
In particular, in Sec.~\ref{sec:verification-cylinder} the limit of $\epsilon \rightarrow 0$ is considered as a verification test, and the oscillation frequency of SAW is investigated in the absence of energetic particles. In Sec.~\ref{sec:TAE} and \ref{sec:EPM}, respectively TAE modes and continuum modes with zero shear are investigated in the presence of energetic particles, and their dynamics is discussed in relation to the continuum topology given by Eq.~\ref{eq:continuum-formula}.

\newpage

\section{Continuous spectrum in a cylinder.}
\label{sec:verification-cylinder}

\subsection{Axisymmetric continuum.}

In this Section, we show the results of the tests of NEMORB for the simplest plasma confinement geometry, i.e. a cylindrical geometry. This is achieved by choosing analytical magnetic equilibria with very small inverse aspect ratio ($\epsilon=0.01$).
Numerical simulations with flat q profiles are performed, where an initial perturbation is let evolve in time (without EP) and the SAW oscillation frequency is measured, for different values of q, electron mass, and electron beta $\beta_e = 8\pi n_e T_e/ B_0^2$ (where $\beta_e$ regulates the density in NEMORB). This frequency is the natural oscillation frequency of the plasma and takes the name of continuous spectrum, or simply continuum~\cite{Chen74}.

Firstly, the continuum for axisymmetric perturbations (corresponding to a toroidal mode number n=0) is considered. We choose an analytical equilibrium with magnetic field on axis $B=2.4$ T, major radius $R=1.667$ m, minor radius $a=0.01667$ m, $\rho^*= \rho_s / a = 1/50$ (with $\rho_s = c_s/\Omega_i$ being the sound gyroradius, and $c_s=\sqrt{T_e/m_i}$ being the sound speed) and $\beta_e=2\cdot 10^{-4}$.
Flat temperature and density profiles are chosen, with the ratio of electron to ion temperature $\tau_e (\rho) = T_e(\rho)/T_i(\rho) = 1$ (for all simulations described in this paper). An axisymmetric ion gyrocenter density perturbation is initialized and let evolve in time. Dirichlet boundary conditions and Neumann boundary conditions are imposed respectively to the external and internal boundaries for the potentials.

\begin{figure}[b!]
\begin{center}
\includegraphics[width=0.45\textwidth]{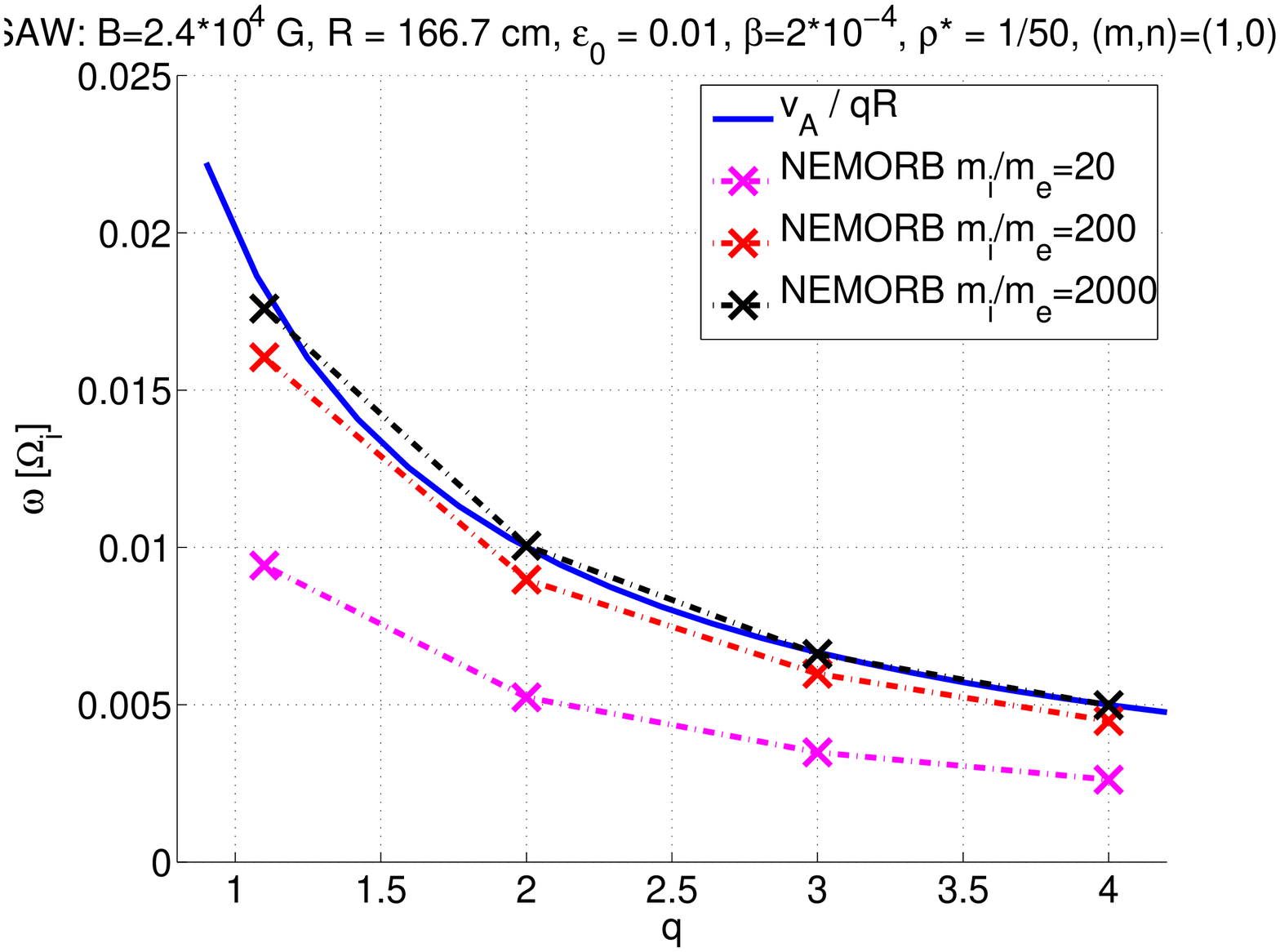}
\includegraphics[width=0.45\textwidth]{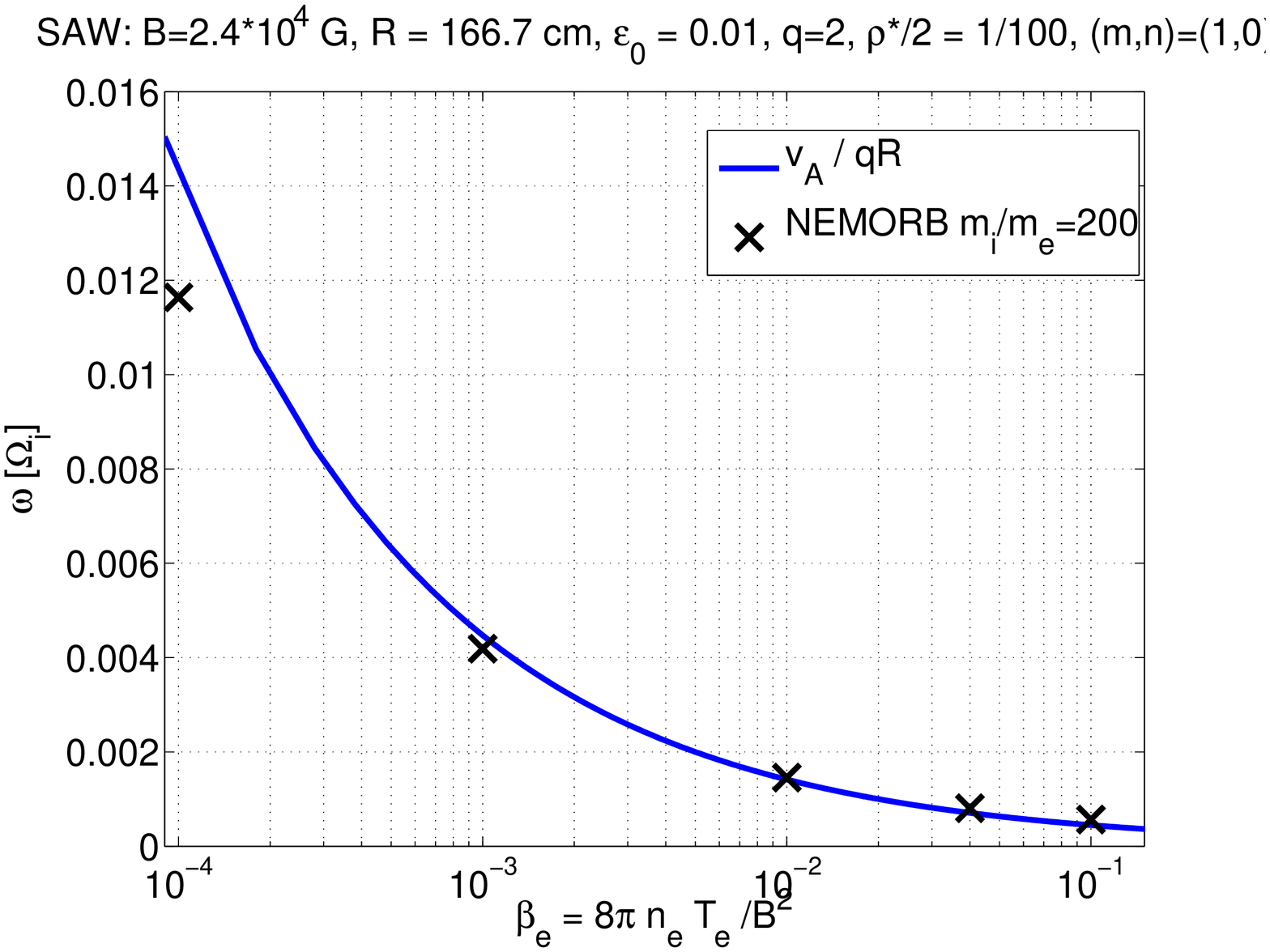}
\caption{On the left, frequency dependence on q, obtained with for different simulations with different value of q (flat q profiles are adopted for each simulations here), and with only m=1, n=0 poloidal component.
Here the cylinder limit ($\epsilon=0.01$) is considered. Convergence with the prediction of ideal MHD theory is found for electrons 200 to 2000 times smaller than ions, for this value of $\rho^*=1/50$ and $\beta_e = 2*10^{-4}$. On the right, a scan of the SAW continuum frequency for different values of the electron beta, and $m_e/m_i = 1/200$. Good stability properties are found, up to values of $\beta_e \sim 0.1$.}\label{fig:cont-axisymm-n0}
\end{center}
\end{figure}

The signal is observed to oscillate and the oscillation frequency is measured. Different simulations with different values of q are performed (see Fig.~\ref{fig:cont-axisymm-n0}), and the scan of the frequency vs q is compared with the ideal MHD prediction for the continuum given by Eq.~\ref{eq:continuum-formula}, which reduces to $\omega_{SAW} = v_A/ q R$ for axisymmetric geometries and axisymmetric oscillations.
Different scans for different values of electron mass are performed, and we note that for this value of $\rho^*$, a good convergence is found for values of electrons mass 2000 times smaller than the ion mass. A scan in $\beta_e$ is also performed, for the case with q=2, and $m_e/m_i = 200$. The code is found to be stable also at this large values of $\beta_e$, where a good convergence with ideal MHD prediction is observed.

\subsection{Non-axisymmetric continuum.}

Similar simulations like in the previous section are described here, but for non-axisymmetric perturbations. The same magnetic equilibrium profile as described in the previous section is considered, with same value of plasma temperature and density. Several simulations are performed, each of them with a different value of q, and each of them with flat q profile.

In this case, we consider the evolution of modes with toroidal mode number n=2 and poloidal mode numbers m=4 and m=5. Like in previous section, no EP population is initialized, therefore our SAW oscillations are stable. We measure the frequency like in the axisymmetric simulations, by measuring the period of oscillation of the potential for simulations where each poloidal component is let evolve in time.

A good match is found, for both branches $m=4$ and $m=5$, with ideal MHD prediction, Eq.~\ref{eq:continuum-formula}, which reduces in the cylinder limit to $\omega_{SAW} = v_A (m-n q) / q R$. Like predicted, in this cylinder limit, no gap in the continuous spectrum due to toroidicity is found at the intersection of the two branches.
A gap forms when toroidal curvature is introduced, by increasing the value of $\epsilon$, and consequently a further degree of nonuniformity breaks the symmetry in the poloidal angle $\theta$ giving rise to the modification of the two cylinder branches.

In the next Section, simulations with a non-negligible toroidal curvature are performed, and the formation of a global mode with frequency lying in the toroidicity-induced gap is shown.

\begin{figure}[b!]
\begin{center}
\includegraphics[width=0.45\textwidth]{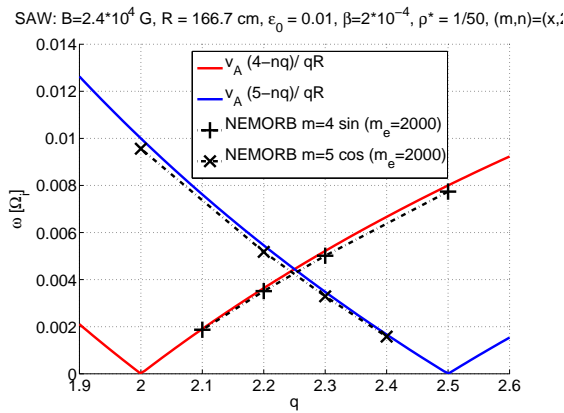}
\caption{Frequency vs q for different simulations with flat q profiles, and with only one poloidal component (m=4 or m=5) evolved in time. Here the cylinder limit ($\epsilon=0.01$) is considered. A good match with the prediction of ideal MHD theory is found for electrons 2000 times smaller than ions, for this value of $\rho^*=1/50$ and $\beta_e = 2*10^{-4}$.}
\end{center}
\end{figure}

\newpage
\section{Toroidicity-induced Alfv\'en Eigenmodes.}
\label{sec:TAE}

\subsection{Equilibrium and numerical setup.}

In this Section, we consider a more realistic value of toroidal curvature, i.e. a larger value of inverse aspect ratio $\epsilon$. In this case, the continuous spectrum gap opens at the intersection of two neighbor cylinder branches m and m+1, and a global eigenmode is created with frequency lying within the continuum gap. This mode takes the name of toroidicity-induced Alfv\'en eigenmode (TAE)~\cite{Cheng85}.

The equilibrium for the simulations shown in this Section has been chosen consistently with the 
ITPA benchmark~\cite{Koenies12}. The equilibrium magnetic field is given by analytical toroidal flux surfaces without Grad-Shafranov shift, and with magnetic field on axis $B=3$ T, major radius $R=10$ m, minor radius $a=1$ m.
The corresponding q-profile is parabolic, with
minimum value of $q(0)=1.72$ and maximum value of $q(1)=1.84$ (a slightly steeper profile was resulting from the VMEC equilibrium code, used for MHD equilibria and described as the reference for gyrokinetic code EUTERPE in the original ITPA publication~\cite{Koenies12}).
Flat temperature and density profiles are considered for bulk ions and electrons, with electron temperature corresponding to a value of $\rho^*= \rho_s / a = 1/927$, $\tau_e=1$, and electron pressure corresponding to a value of $\beta_e = 9.1\cdot 10^{-4}$. This choice of the plasma temperature and density is made in order to have TAEs with frequency well approximated by ideal MHD (the effects of plasma compressibility turn out to be negligible). Electrons 200 times lighter than ions are considered.

A distribution function Maxwellian in $v_\|$ is considered for the EP population. The EP averaged concentration is $<n_{EP}>/n_e = 0.00307$ with radial profile given by:
\begin{equation}\label{eq:TAE-n_EP}
n_{EP}(s)/n_{EP}(s_0) = \exp [-\Delta \, \kappa_n \tanh ((s-s_0)/\Delta)] 
\end{equation}
with $s_0=0.5$, $\Delta=0.2$, and $\kappa_n = 3.333$. In the simulations shown in this paper, due to the very small amount of fast particles, the bulk ion and electron profiles are not corrected to satisfy quasi-neutrality to a higher level of accuracy.

\begin{figure}[b!]
\begin{center}
\includegraphics[width=0.45\textwidth]{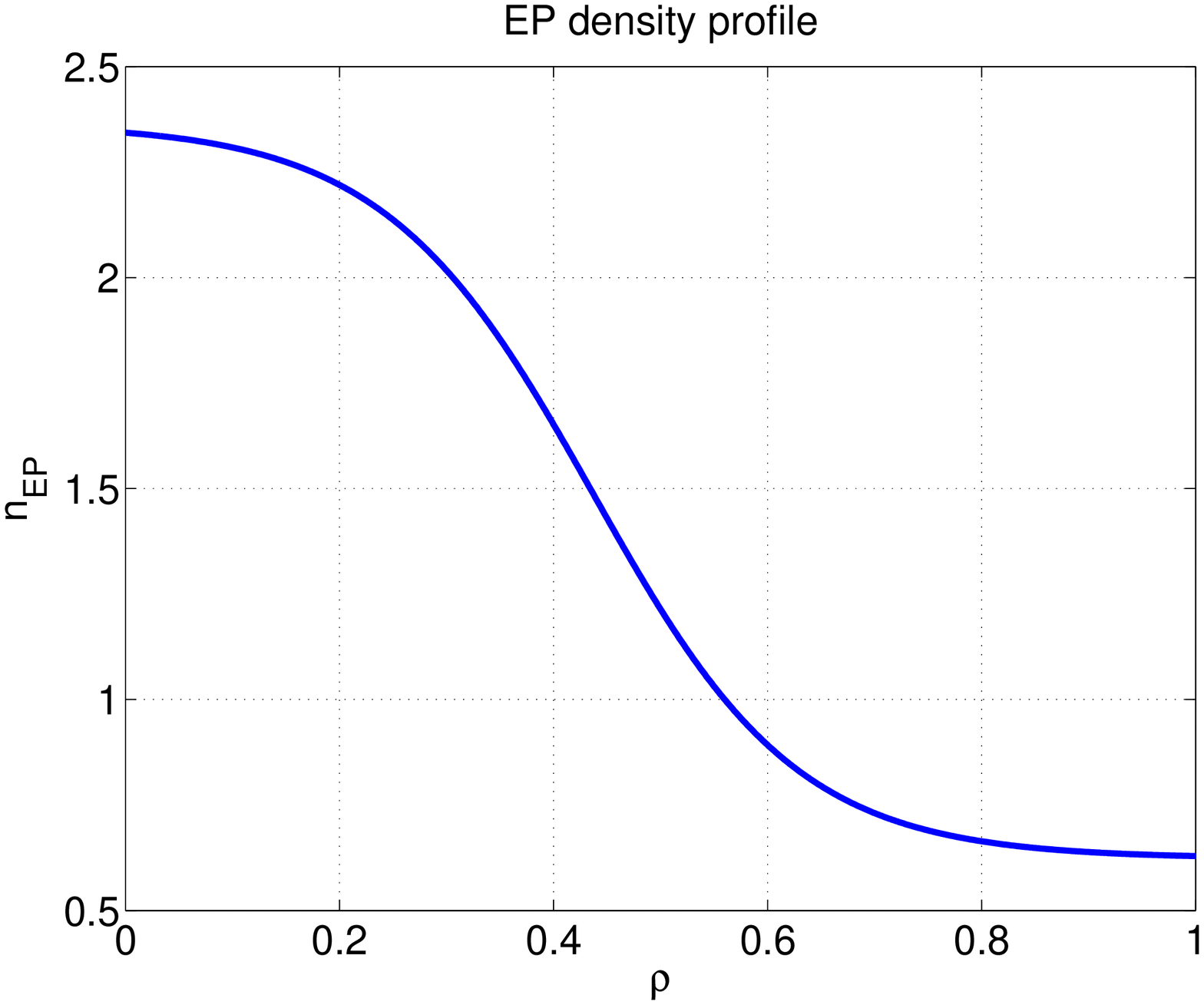}
\includegraphics[width=0.45\textwidth]{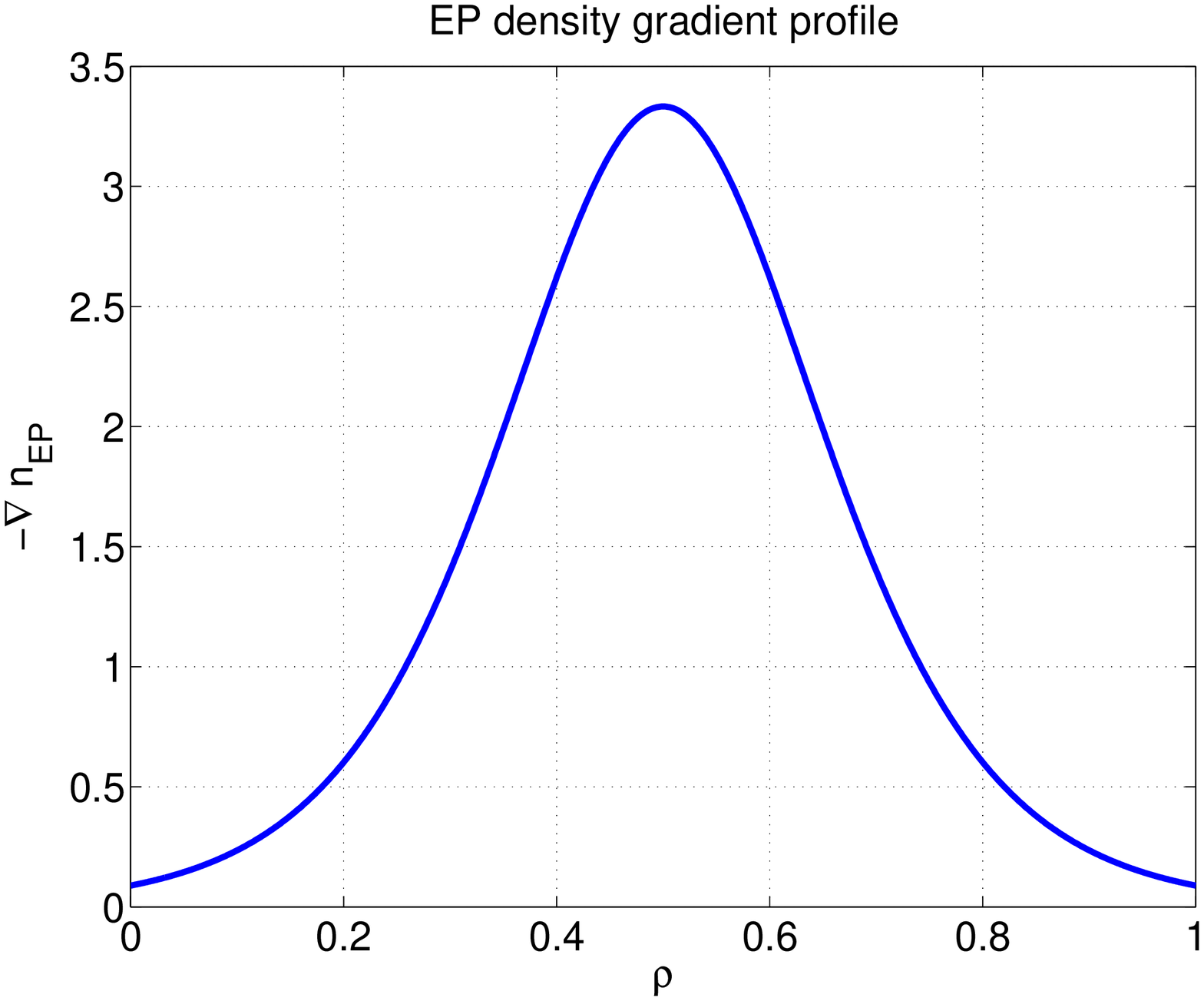}
\vskip -1em
\caption{On the left, energetic particle density profile, like in Eq.~\ref{eq:TAE-n_EP}, normalized in volume. On the right, corresponding logarithmic gradient normalized with the minor radius $a$.}\label{fig:TAE-evolution}
\end{center}
\end{figure}

Linear simulations with $10^7$ markers per species have been run, with a space resolution of (ns,nchi,nphi)=(256,256,64), and a time step of dt=20 $\Omega_i^{-1}$. A filter in mode numbers is applied, which keeps only n=6 mode, and m=9,10,11,12.
A radial domain going from $s=0.1$ to $s=0.9$ is considered for the evolution of the potentials. No finite-Larmor-radius effects are studied in this paper, meaning that the gyroaverage operators are set to 1 for all species in the code, for the simulations described here.

\subsection{TAE frequency and structure.}

We initialize a perturbation in the density of the ion gyrocenters at t=0, with n=6 and m=10 and 11, and with a radial structure calculated in order to give a perturbation of potential radially localized around s=0.5 at t=0.
The perturbed vector potential is measured at different radii as a function of time, at a given poloidal angle. The frequency of the mode is found to depend on the EP concentration (see Fig.~\ref{fig:TAE-evolution}, where a comparison with the continuum X-point is shown).
The limit for vanishing EP concentration matches well with the center of the TAE gap calculated with Eq.~\ref{eq:continuum-formula}.

\vskip -1em

\begin{figure}[h!]
\begin{center}
\includegraphics[width=0.42\textwidth]{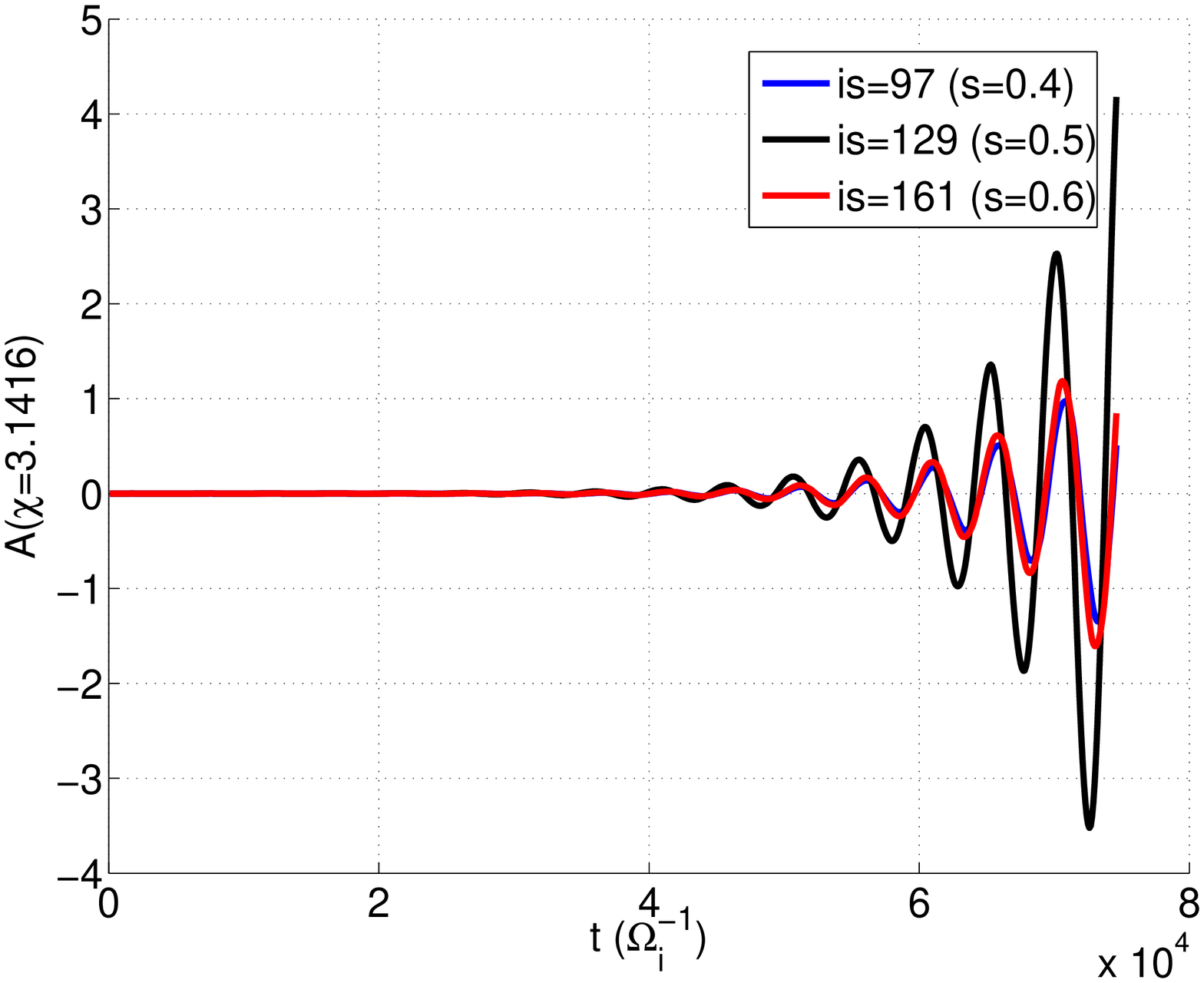}
\includegraphics[width=0.4\textwidth]{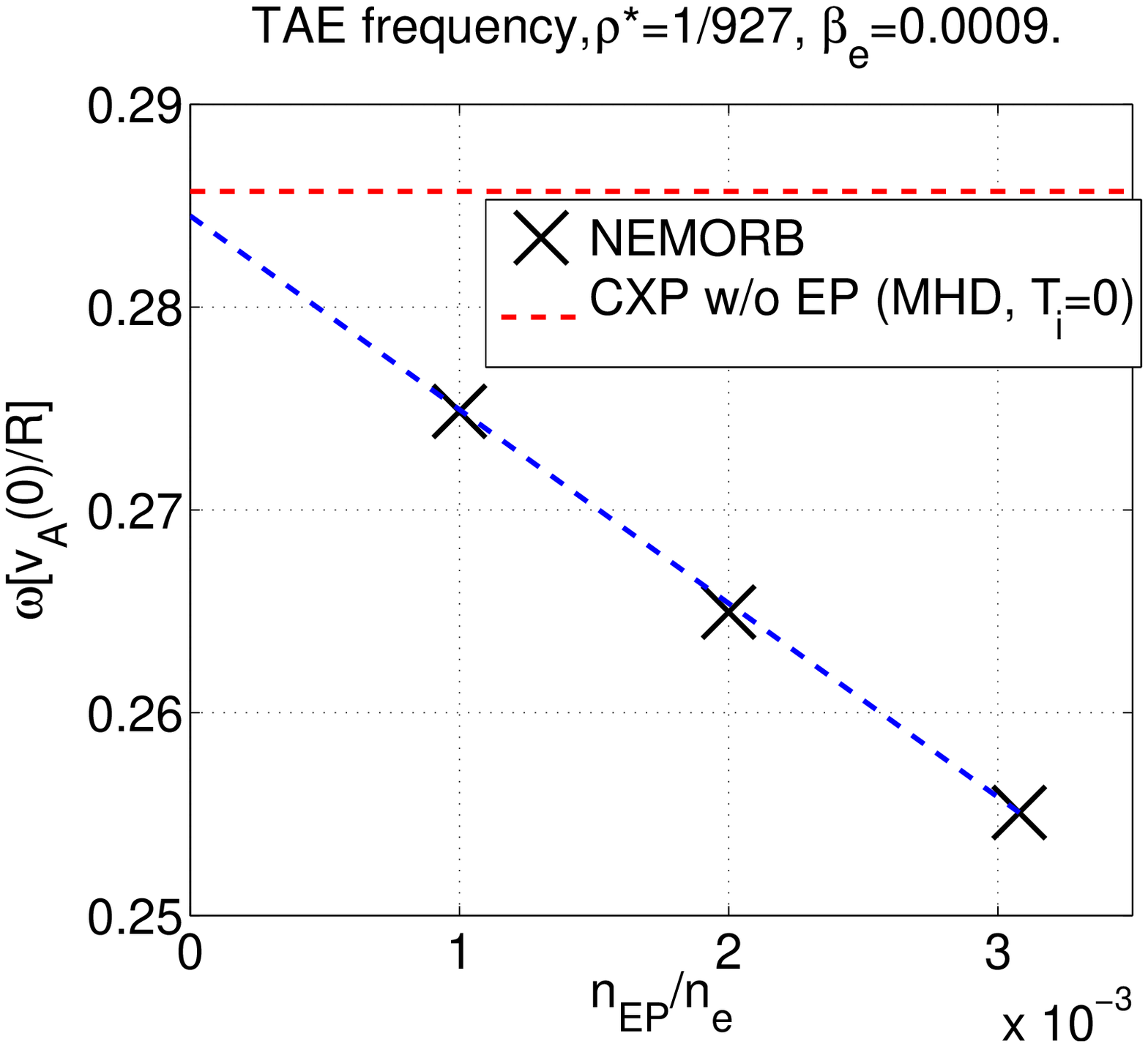}
\vskip -0.5em
\caption{On the left, evolution in time for the case with $n_{EP}/n_e$ = 0.003, $T_{EP}$ = 500 keV. On the right, dependence of the TAE frequency on the EP averaged concentration.}\label{fig:TAE-evolution}
\end{center}
\end{figure}

\vskip -1em

\begin{figure}[h!]
\begin{center}
\includegraphics[width=0.45\textwidth]{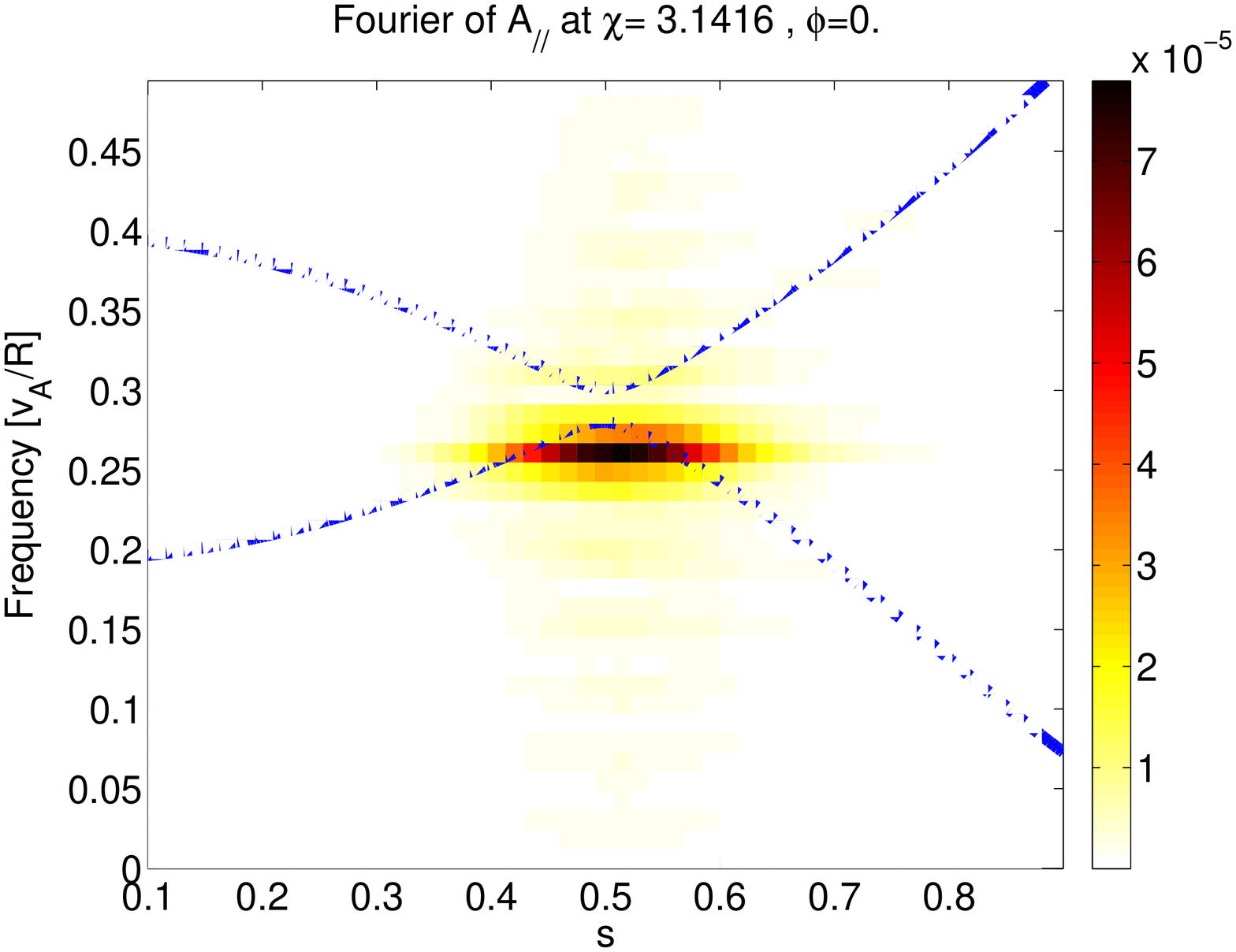}
\includegraphics[width=0.47\textwidth]{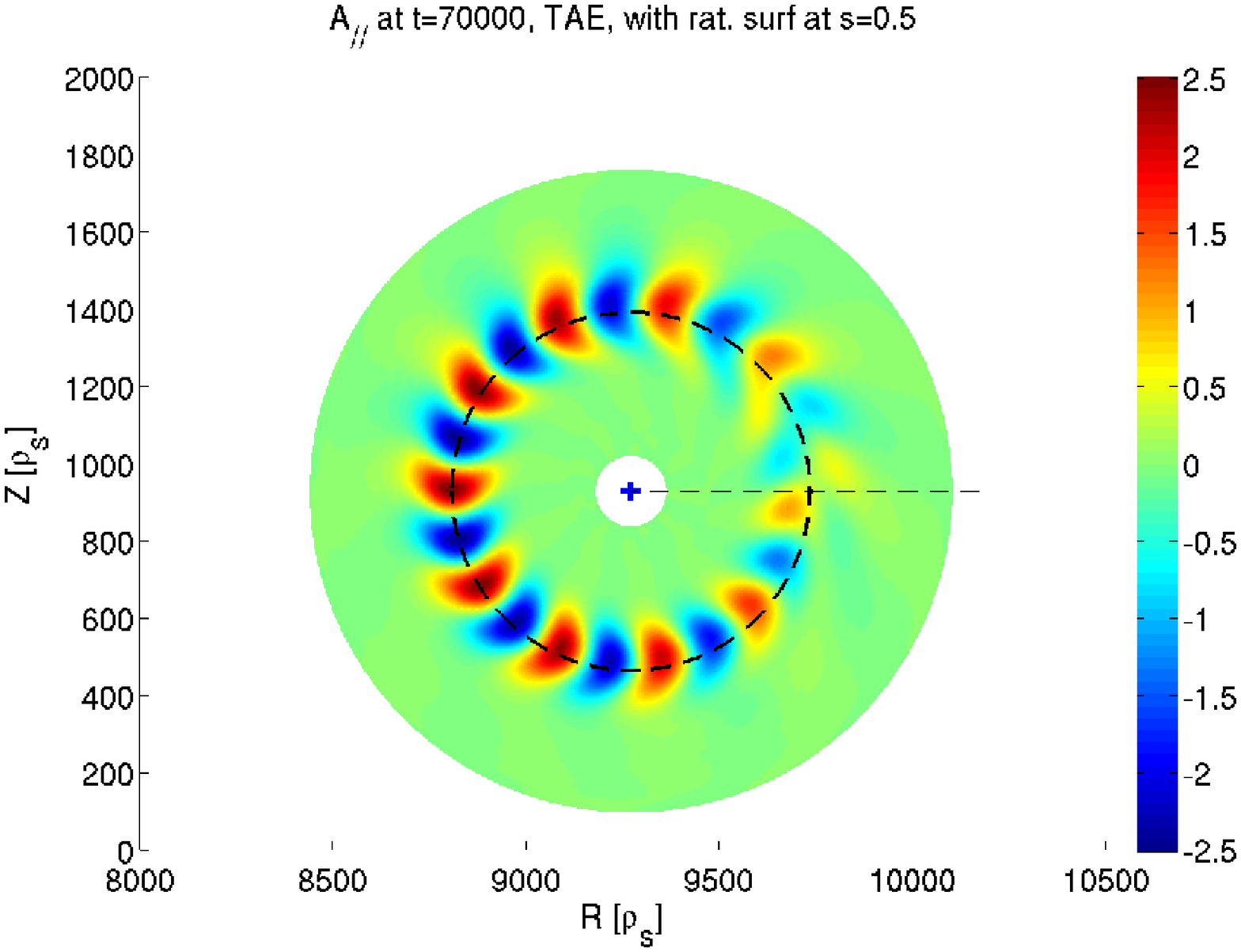}
\caption{On the left, Fourier transform in time of the vector potential for the case with $n_{EP}/n_e$ = 0.003, $T_{EP}$ = 500 keV. On the right, structure of the vector potential at t=60000 $\Omega_i^{-1}$.}\label{fig:TAE-structure}
\vskip -0.5em
\end{center}
\end{figure}

\newpage

The mode is found to be peaked radially at the position of the center of the SAW continuum gap, s=0.5, and has poloidal structure of the vector potential with characteristic ballooning features (i.e. mixed m and m+1 features) at the high-field side (see Fig.~\ref{fig:TAE-structure}). At the high-field side, around the s=0.5 flux surface, the mode shows a phase shift in $\theta$ with respect to the value at s=0.5.
This creates a ``boomerang'' shape, which will be studied in details for continuum modes in Sec.~\ref{sec:EPM}.

\subsection{TAE growth rate.}

\begin{figure}[h!]
\begin{center}
\includegraphics[width=0.47\textwidth]{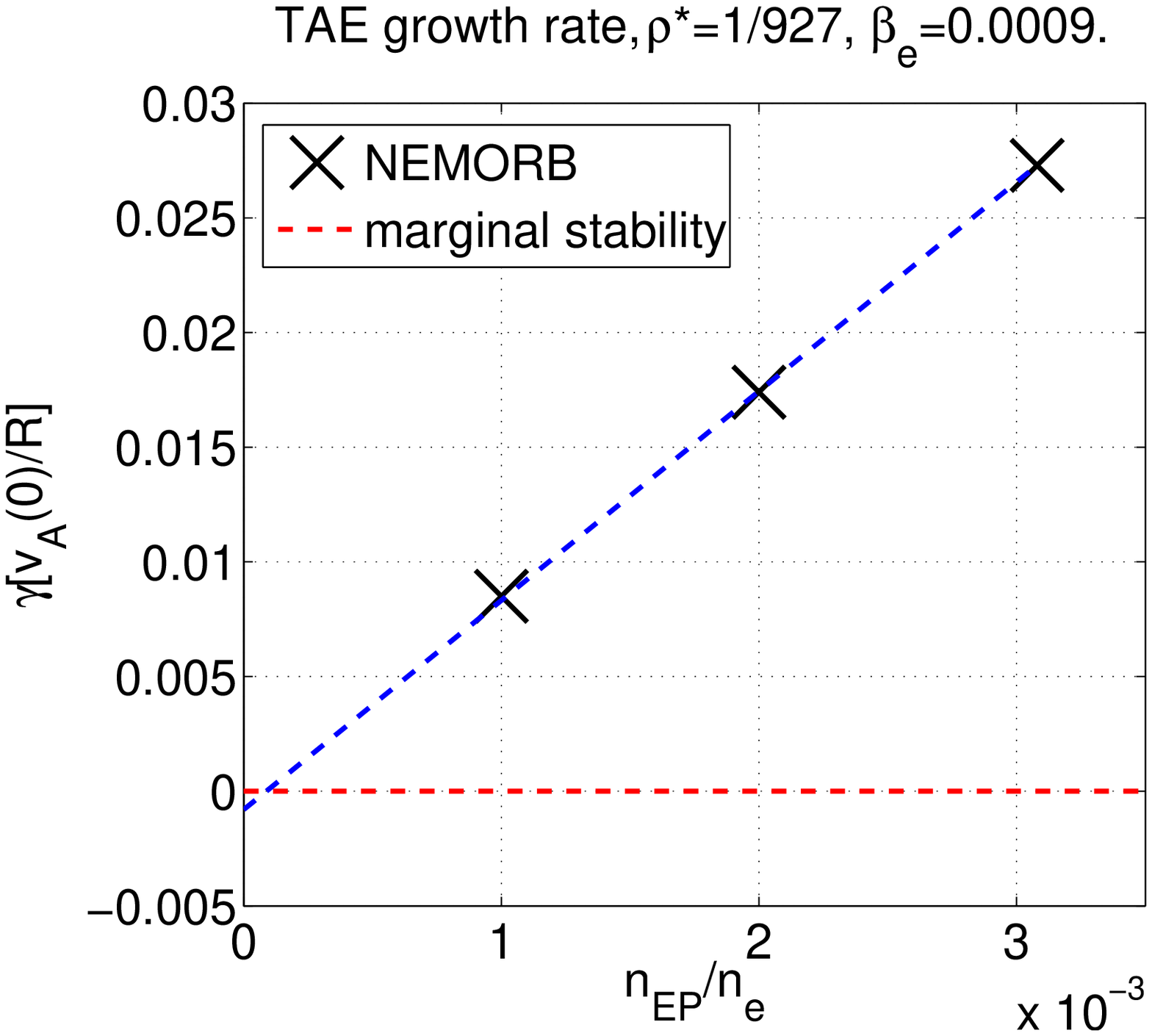}
\includegraphics[width=0.48\textwidth]{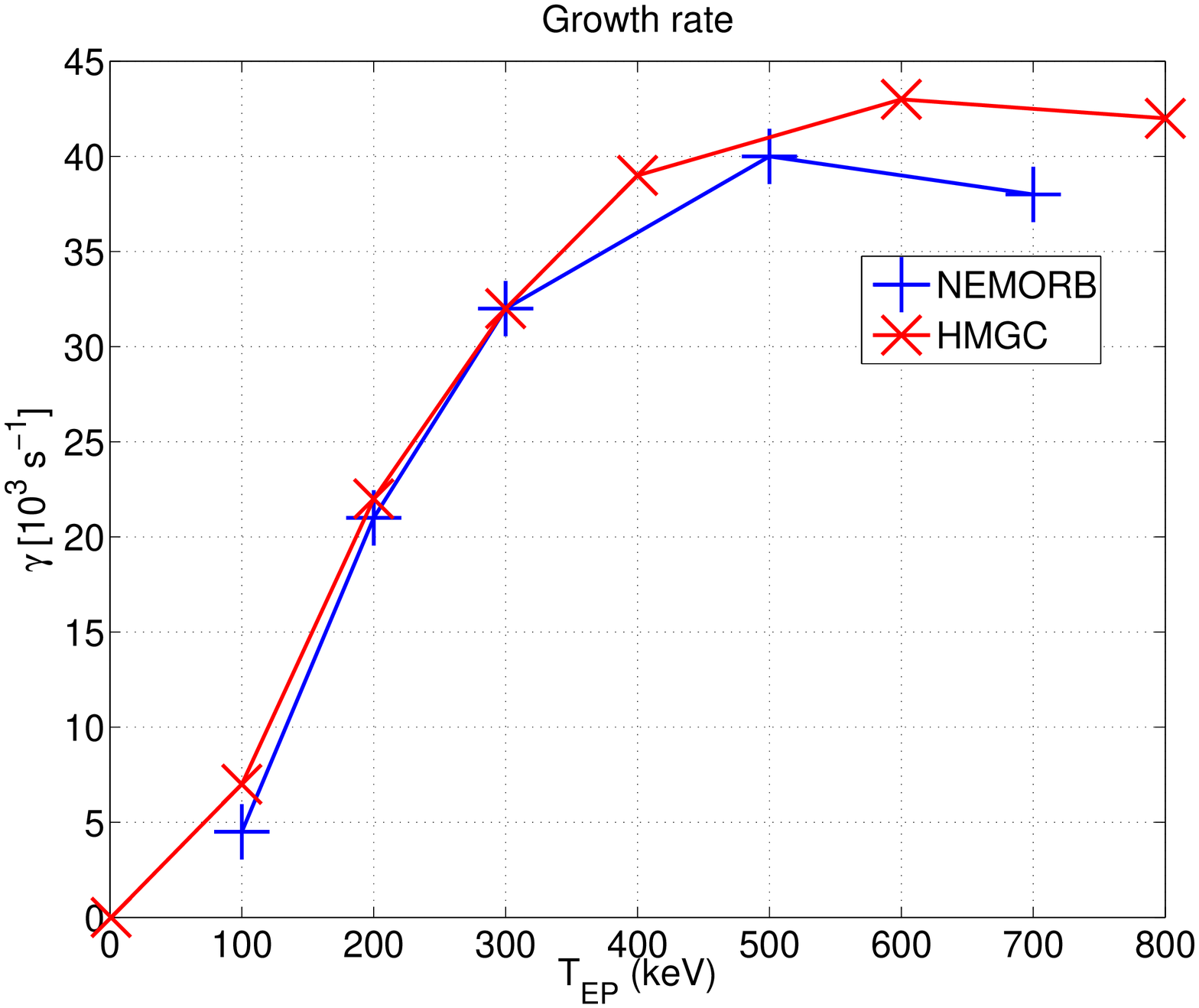}
\caption{On the left, dependence of the TAE growth rate on the EP average concentration, for $T_{EP}=500$ keV. On the right, comparison of the dependence of the growth rate on the EP temperature, for NEMORB and HMGC. HMGC data are taken from Ref.~\cite{Koenies12}.}\label{fig:TAE-growth-rate}
\end{center}
\end{figure}

Due to the EP radial density gradient, energy is pumped from the EP thermal energy to the macroscopic SAW kinetic energy. This leads to an exponential growing of the mode amplitude. The dependence of the growth rate on the EP concentration is found to be linear (see Fig.~\ref{fig:TAE-growth-rate}).
The damping is found to be very small for this tokamak configuration, confirming that we are deep into the MHD regime. The drive dependence on the EP temperature is also compared with results of the hybrid MHD-gyrokinetic code HMGC shown in Ref.~\cite{Koenies12}, where the drive is estimated by adding the measured growth rate and the damping rate (evaluated, for each EP temperature, as in Fig.~\ref{fig:TAE-growth-rate}-left, and negligible in NEMORB case).
This comparison can be seen in Fig.~\ref{fig:TAE-growth-rate}-right. A very good agreement is found, except at very high temperatures, where a difference is present between the results of the two codes.


\newpage
\section{Energetic-particle driven continuum modes.}
\label{sec:EPM}

\subsection{Equilibrium and numerical setup.}

In this Section, we want to study how EP drive Alfv\'en modes of the continuum unstable. We focus on modes where no continuum damping occurs. This is done by driving continuum modes centered radially where the gradient of the continuous spectrum vanishes. 
We consider two equilibria, one where the modes are excited at the continuum accumulation point (CAP) created by an inversion of the magnetic shear, and another one where the continuum frequency is constant in radius.
In the presence of a reversed shear, AEs can exist which are usually referred to as reversed-shear-induced AE (RSAE)~\cite{Breizman05}. Nevertheless, in our case, due to the small value of $\epsilon$, the mode frequency in the limit of zero EP concentration tends to the CAP (and not to a discrete frequency lying in the continuum gap, far from the CAP), therefore we refer to them as EP driven continuum modes (EPM)~\cite{Chen07}.

We consider the same analytical equilibrium as in Section~\ref{sec:TAE} (same minor and major radius, concentric flux surfaces with no Shafranov shift), with same magnetic field intensity at the axis (B=3T), but with a different poloidal component, yielding a different safety factor profile. The safety factor has a value of 1.85 at the axis, it decreases from $\rho$=0 to $\rho$=0.5, where the minimum value is located ($q(\rho=0.5)$=1.78), and then it raises to the edge, where it reaches the maximum value ($q(\rho=1)$=2.6).
Ion and electron density and temperature profiles are flat. We consider two regimes with ion temperature respectively $T_i$= 3.44 keV (corresponding to $\rho^*=1/500$) and $T_i$=55.1 keV (corresponding to $\rho^*=1/125$). The electron to ion temperature ratio is always kept at 1 throughout this paper, $\tau_e=1$, and the electron pressure is chosen for a value of $\beta_e = 5\cdot 10^{-4}$.


\begin{figure}[h!]
\begin{center}
\includegraphics[width=0.4\textwidth]{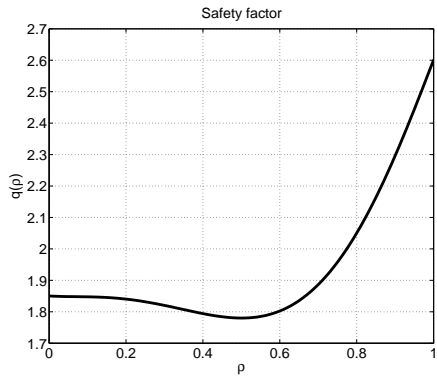}
\vskip -0.5em
\caption{Safety factor profile for the reversed-shear equilibrium.}
\label{fig:EPM-q}
\end{center}
\end{figure}

%

A similar distribution function as in the previous section is considered here for the EP, Maxwellian in $v_\|$ and with a radial density gradient peaked around $s=0.5$. The EP concentration has a radial profile given by Eq.~\ref{eq:TAE-n_EP}, with $s_0=0.5$, $\Delta=0.16$, and $\kappa_n = 10$. The EP temperature is kept fixed for all simulations of this section, with $T_{EP} = 5510 keV$.
No finite-Larmor-radius effects are considered here (like in the previous section), i.e. the gyroaverage operators are set to their value of $k_\perp \rho_{i,e,f}$ = 0 (for the bulk-ion, electron and fast species).
Modes with n=6, m=11 and modes with n=6, m=10 are considered. The characteristic number of markers for bulk ions, electrons and fast ions is $10^6$. Electrons 2000 times lighter than ions are chosen. The characteristic space and time resolutions are (ns,nchi,nphi)=(256,256,64), dt=5  $\Omega_i^{-1}$.

\subsection{EPM frequency and growth rate.}

\begin{figure}[b!]
\begin{center}
\includegraphics[width=0.58\textwidth]{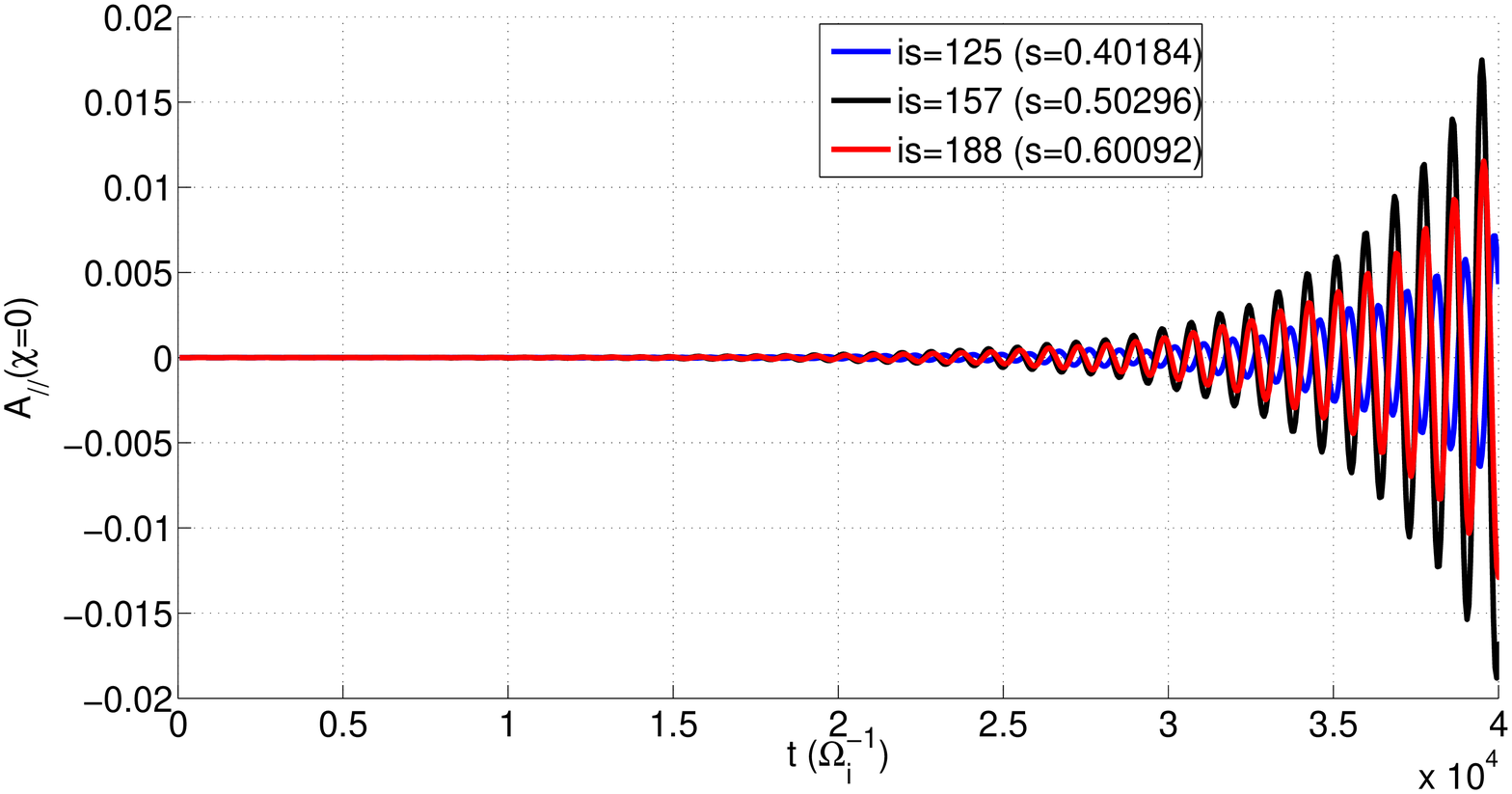}
\includegraphics[width=0.4\textwidth]{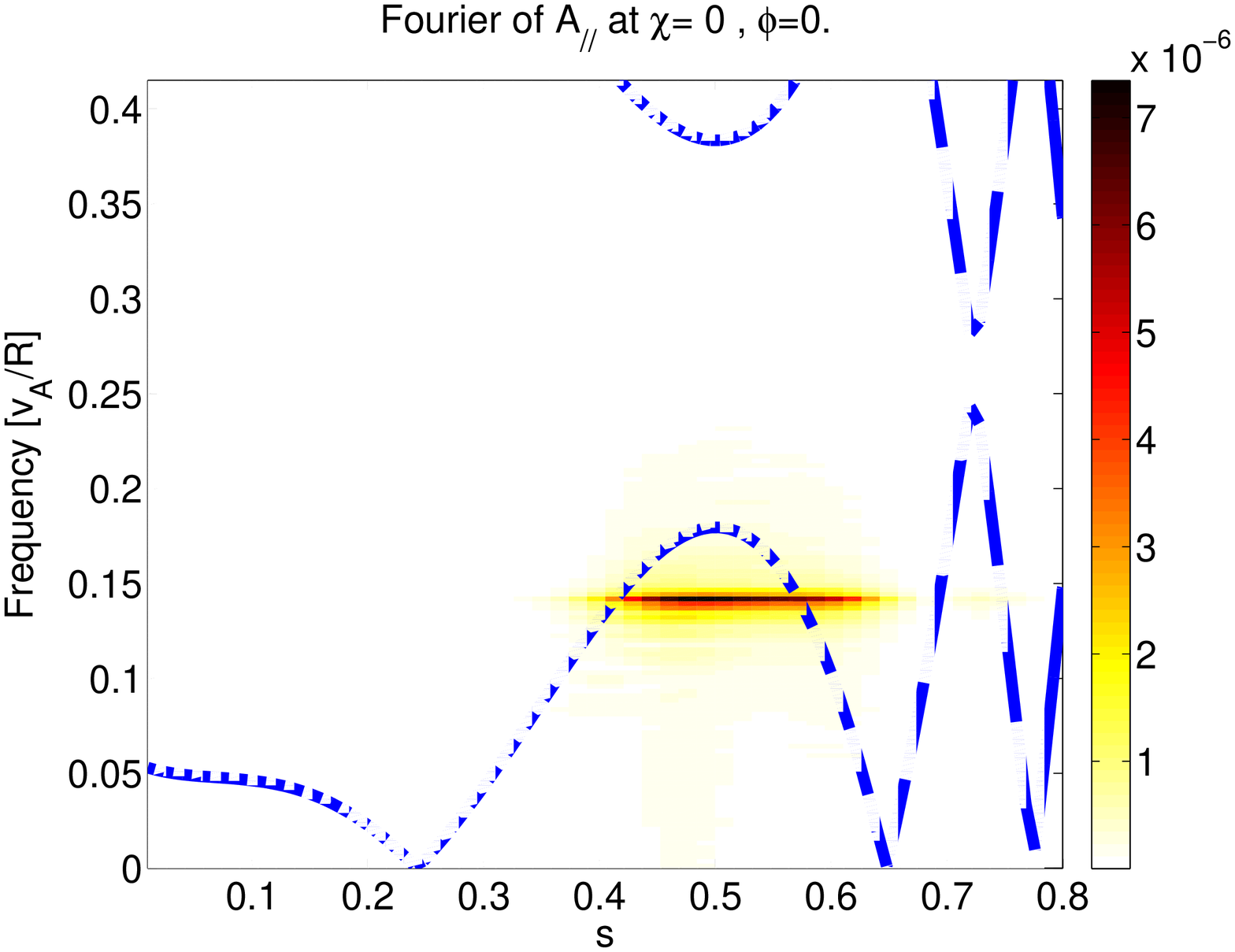}
\vskip -1em
\caption{On the left, $A_\|$ measured at 3 different radial locations, and for a given poloidal angle, for $T_i$=55.1 keV and $n_{EP}/n_e=0.03$, for the reversed-shear equilibrium. On the right, corresponding Fourier transform in time (in red) compared with the theoretical continuum (in blue).}\label{fig:EPM-Fourier}
\end{center}
\end{figure}

For the reversed-shear equilibrium, shear Alfv\'en modes are initialized and observed to grow at the CAP location ($\rho=0.5$), where the EP density gradient is peaked, with a poloidal mode number selected with a filter. Modes with m=10 and modes with m=11 are studied separately. The signal of $A_\|$ is measured at different radial locations, and the Fourier transform in time is performed to calculate the frequency for each radial location.
The mode frequency is compared with the continuous spectrum formula, Eq.~\ref{eq:continuum-formula}, which is crucial to characterize the nature of the unstable mode. As an example, in Fig.~\ref{fig:EPM-Fourier} a mode with filter at m=11 is shown, and the frequency is found indeed near the m=11 continuum branch. Note that the eigenmode structure is observed (the frequency is independent of the radial position). Note also that the frequency lies below the CAP, and not at the CAP frequency, for this EP concentration.

\begin{figure}[t!]
\begin{center}
\includegraphics[width=0.41\textwidth]{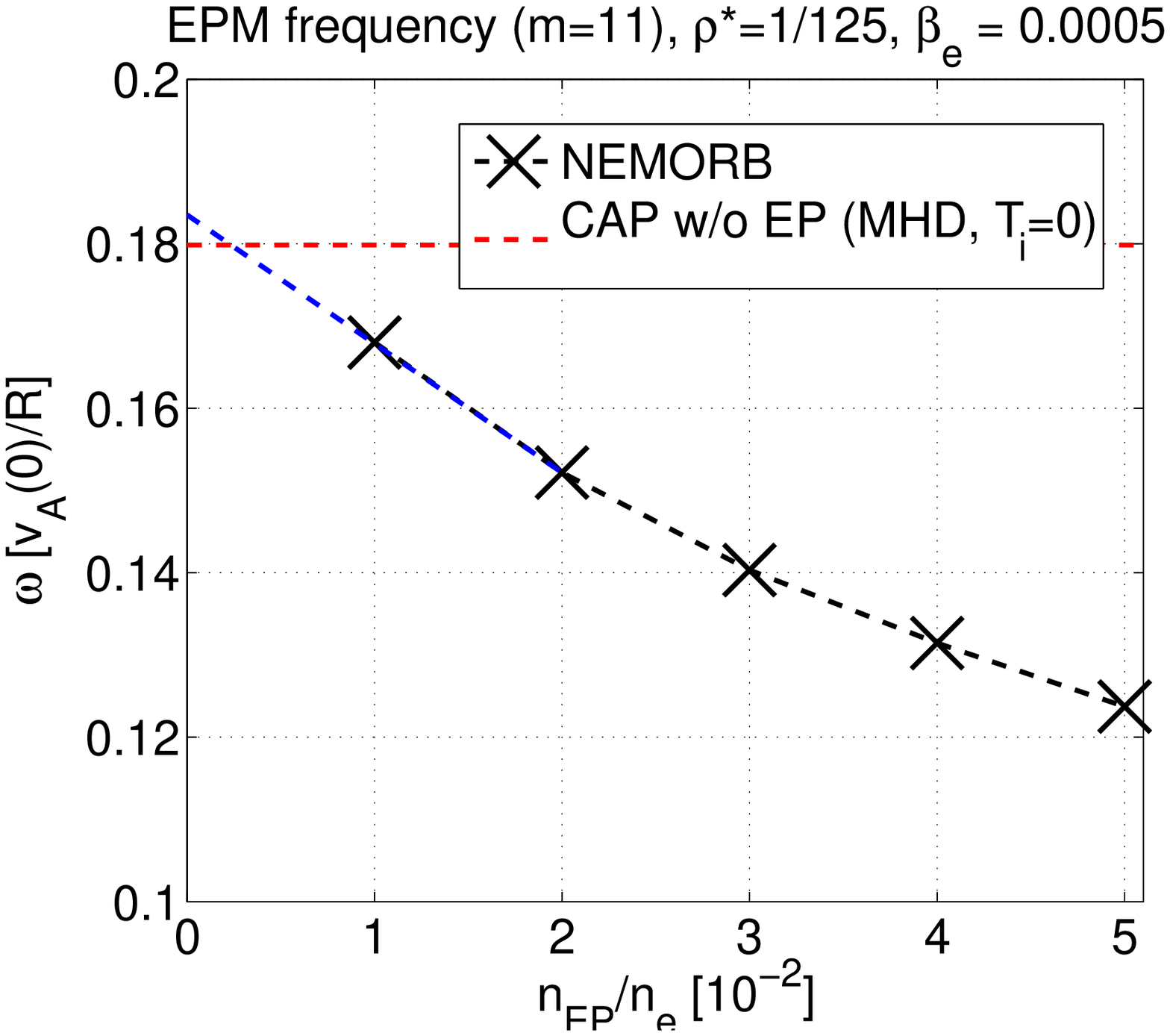}
\includegraphics[width=0.42\textwidth]{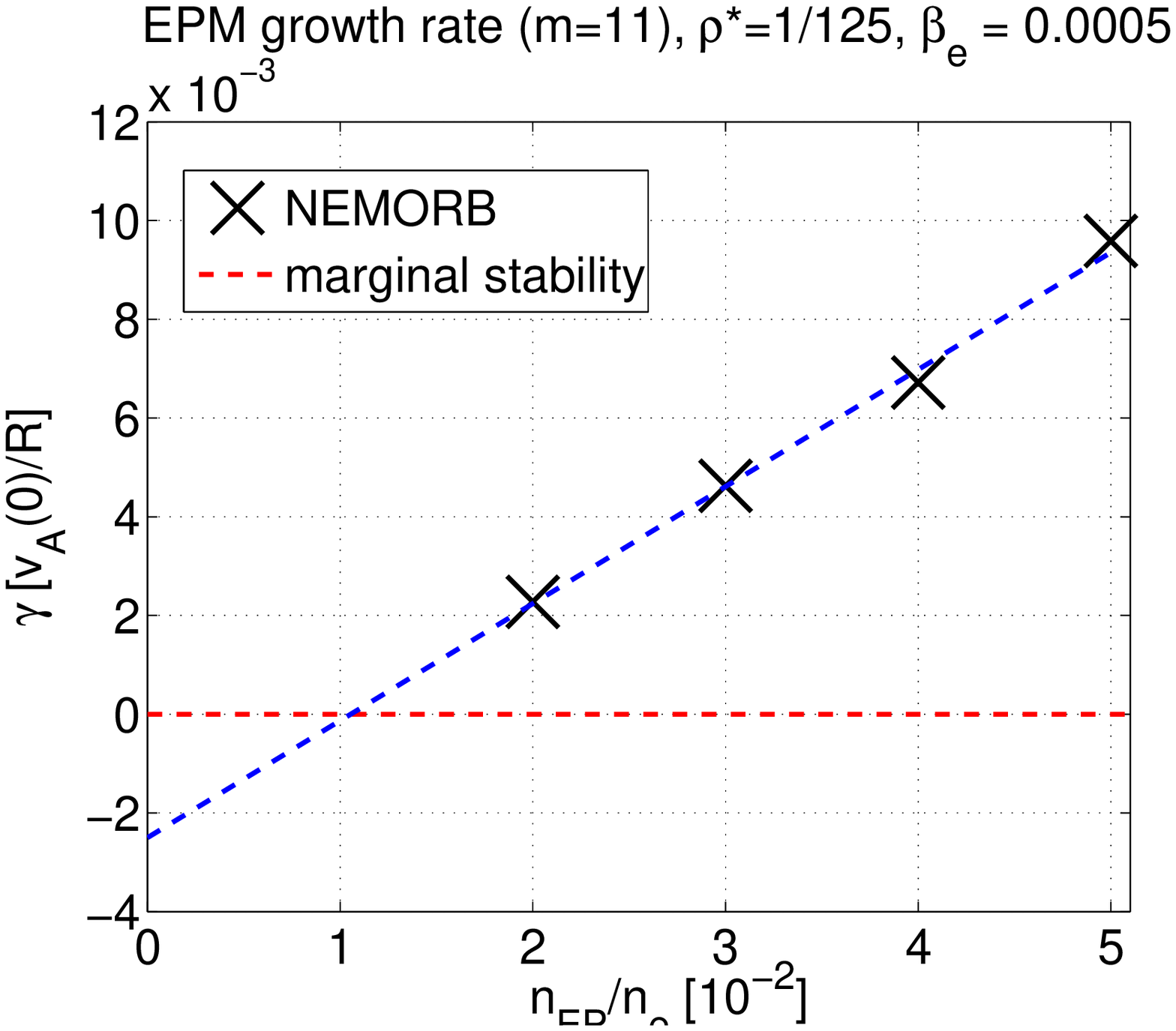}
\vspace{-1em}
\caption{Frequency (left) and growth rate (right) of the m=11 mode for different EP concentrations.
\vspace{-2em}}\label{fig:EPM-omegagamma_nEP-m11}
\end{center}
\end{figure}

\begin{figure}[t]
\begin{center}
\includegraphics[width=0.42\textwidth]{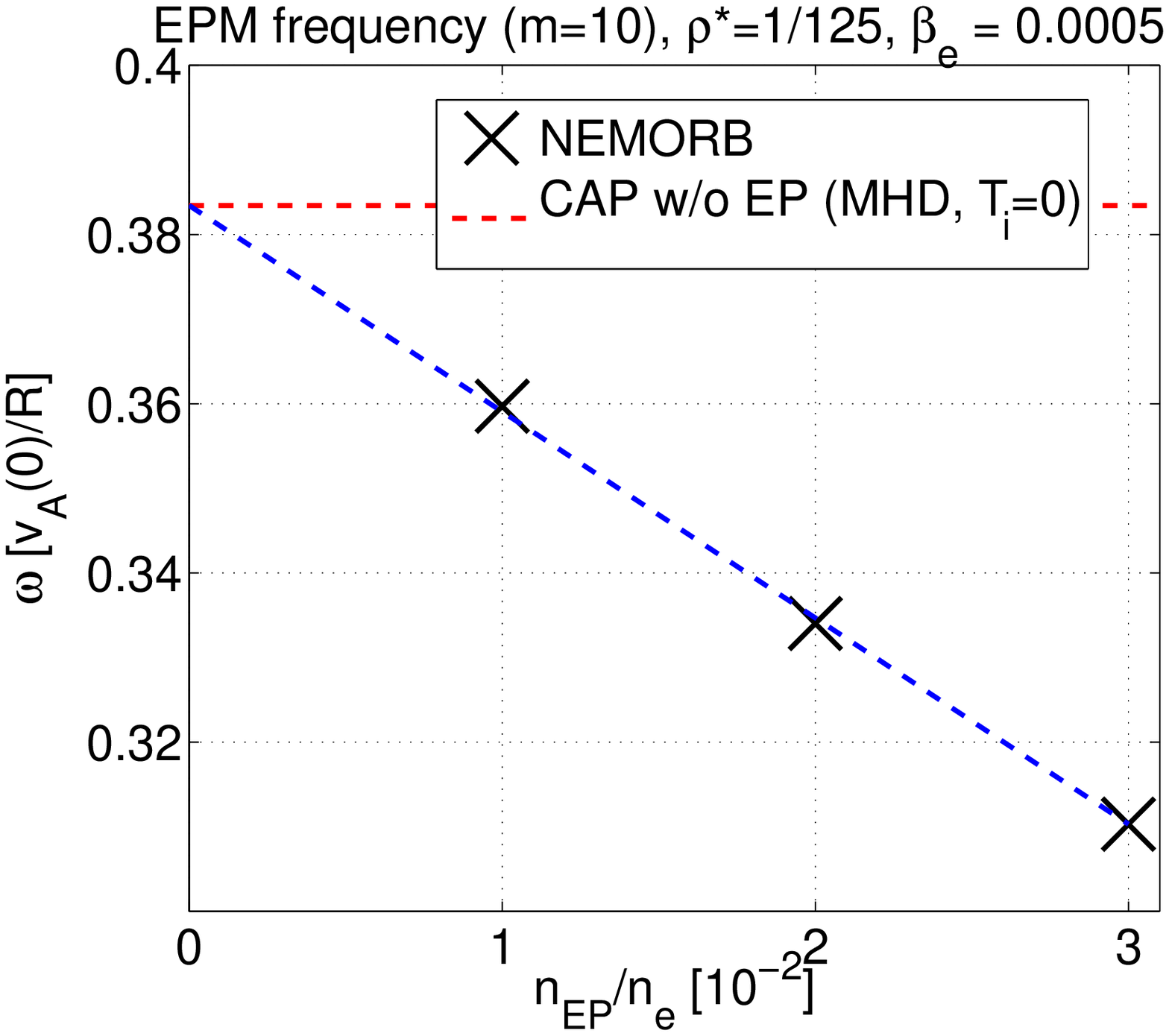}
\includegraphics[width=0.42\textwidth]{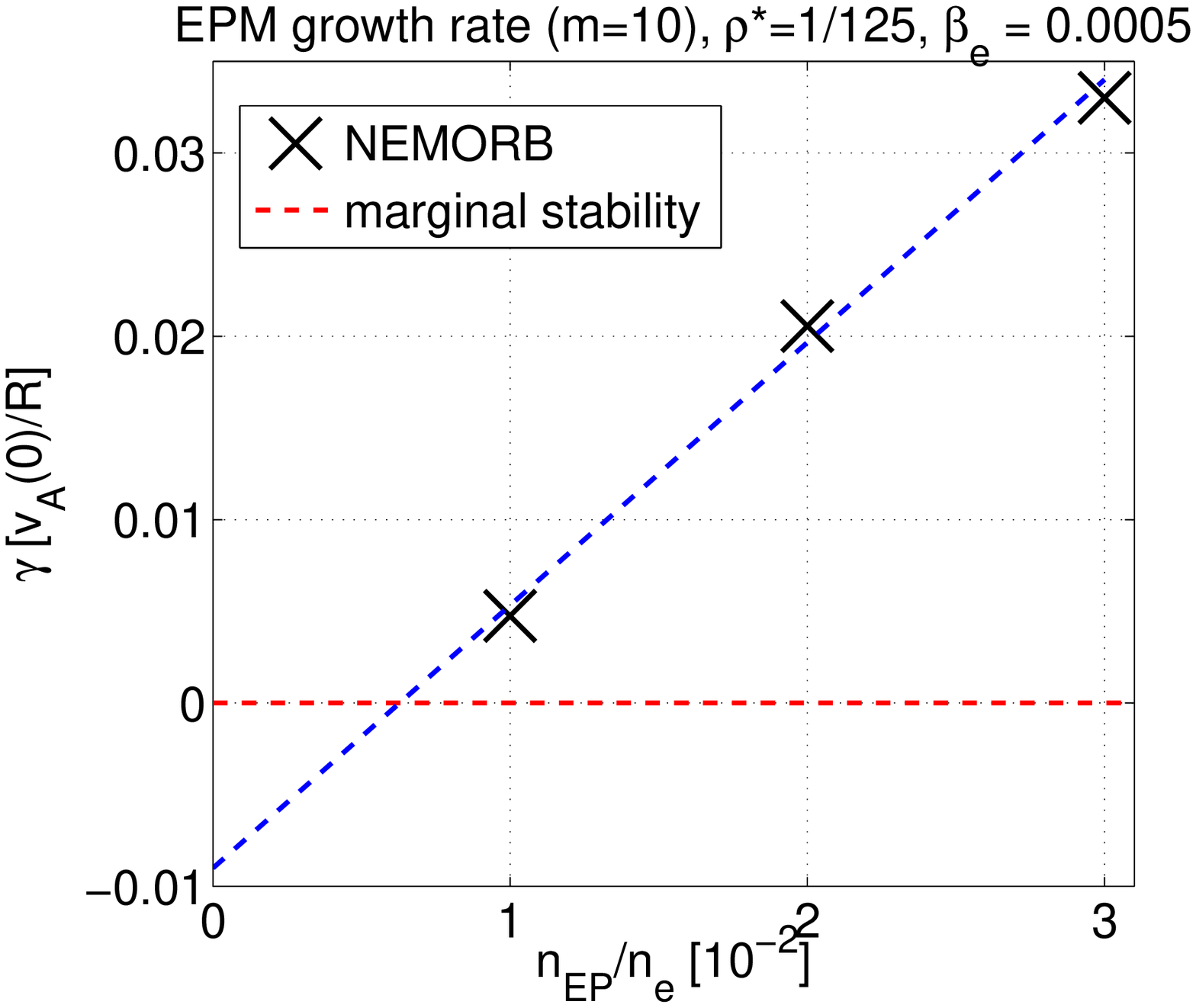}
\vspace{-1em}
\caption{Frequency (left) and growth rate (right) of the m=10 mode, for different EP concentrations.\vspace{-2em}}\label{fig:EPM-omegagamma_nEP-m10}
\end{center}
\end{figure}

The dependence of the mode frequency and growth rate on the EP concentration is shown in Fig.~\ref{fig:EPM-omegagamma_nEP-m11} and Fig.~\ref{fig:EPM-omegagamma_nEP-m10} for the m=11 and m=10 cases. In the same figures, the value of the CAP frequency calculated with Eq.~\ref{eq:continuum-formula} is reported.
The inclusion of plasma compressibility effects can be calculated as in Ref.~\cite{Zonca96}, and gives an upshift of about 0.005 $v_A/R$ for the m=11 mode, and negligible for the m=10 mode. 
We can see that, for both m=10 and m=11 modes, the effect of EP in this regime is to decrease the frequency of the mode. In both cases, the mode frequency tends to the CAP in the limit of zero EP concentration. A comparison with the continuous spectrum topology (Fig.~\ref{fig:EPM-Fourier}), tells us that the m=11 mode enters the continuum in the 
presence of EP, whereas the m=10 mode leaves the continuum and stays into the gap. This difference reflects in a difference in the mode structure, as described in the next section.

The growth rate has also been measured for the m=11 and m=10 modes, for different EP average concentrations (see Fig.~\ref{fig:EPM-omegagamma_nEP-m11} and Fig.~\ref{fig:EPM-omegagamma_nEP-m10}). Except for very high values of growth rates, a linear dependence is found in the EP density. The instability threshold is observed to be lower for the m=10 mode, consistently with the lower damping.

\newpage

\subsection{EPM structure.}

The poloidal structure of $A_\|$ has also been studied for the modes with m=11 and m=10. The poloidal mode number is defined for both cases by means of a filter. The radial structure is centered near $\rho=0.5$ (corresponding to s=$\sqrt{\psi/\psi(1)}=0.54$), where the peak of the EP gradient is located, chosen at the SAW CAP. The external boundary condition is set here at $s=0.8$, in order to isolate the dynamics of interest around $s=0.5$, from TAE modes getting unstable at outer radii for the case where the Landau damping is lower (i.e. for low values of $\rho^*$).

\begin{figure}[b!]
\begin{center}
\includegraphics[width=0.42\textwidth]{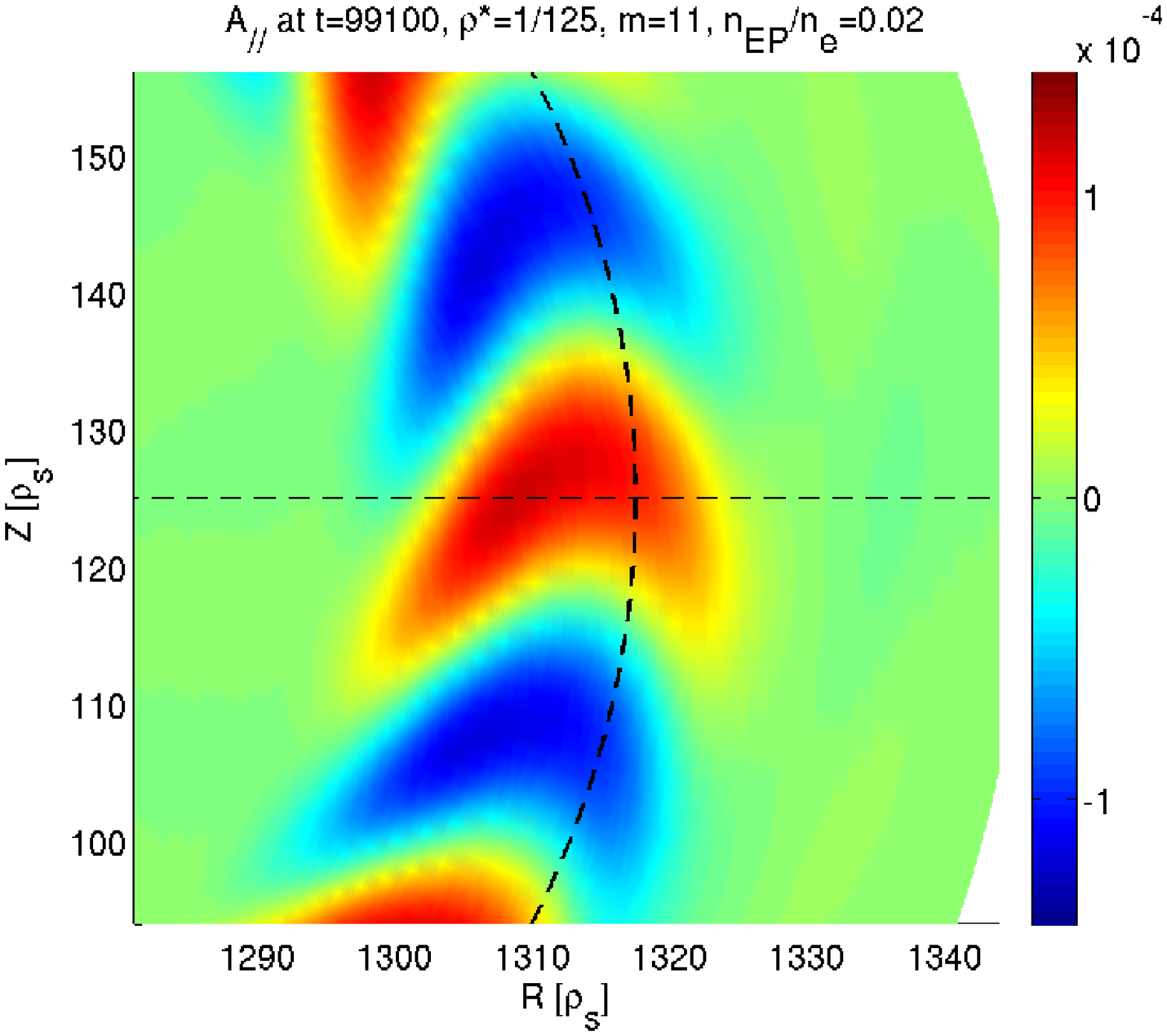}
\includegraphics[width=0.42\textwidth]{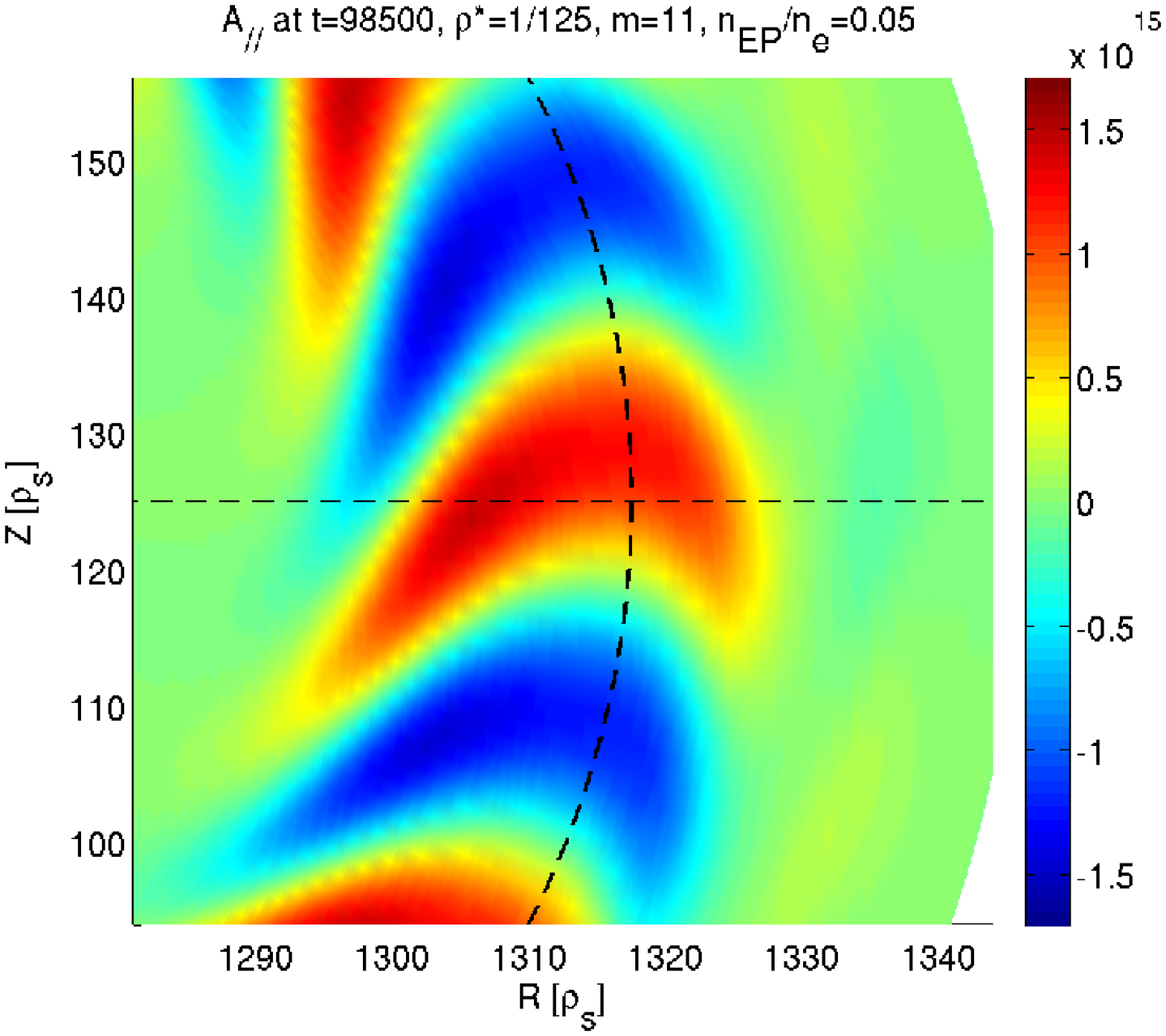}
\vskip -1.4em
\caption{Poloidal structure of $A_\|$ for the m=11 mode, for the case with reversed-shear q profile shown in Fig.~\ref{fig:EPM-q}, $T_i$=55.1 keV ($\rho^*=1/125$). Energetic particles are characterized by $T_{EP}$=5510 keV and respectively $n_{EP}/n_e =0.02$ (left) and  $n_{EP}/n_e =0.05$ (right). Radial mode extension is found to depend on the EP concentration, but no difference is found in the poloidal mode section geometry.}\label{fig:EPMm11-structure-vs_nEP}
\vskip -1em
\end{center}
\end{figure}

\begin{figure}[b!]
\begin{center}\label{fig:EPM-structure}
\includegraphics[width=0.42\textwidth]{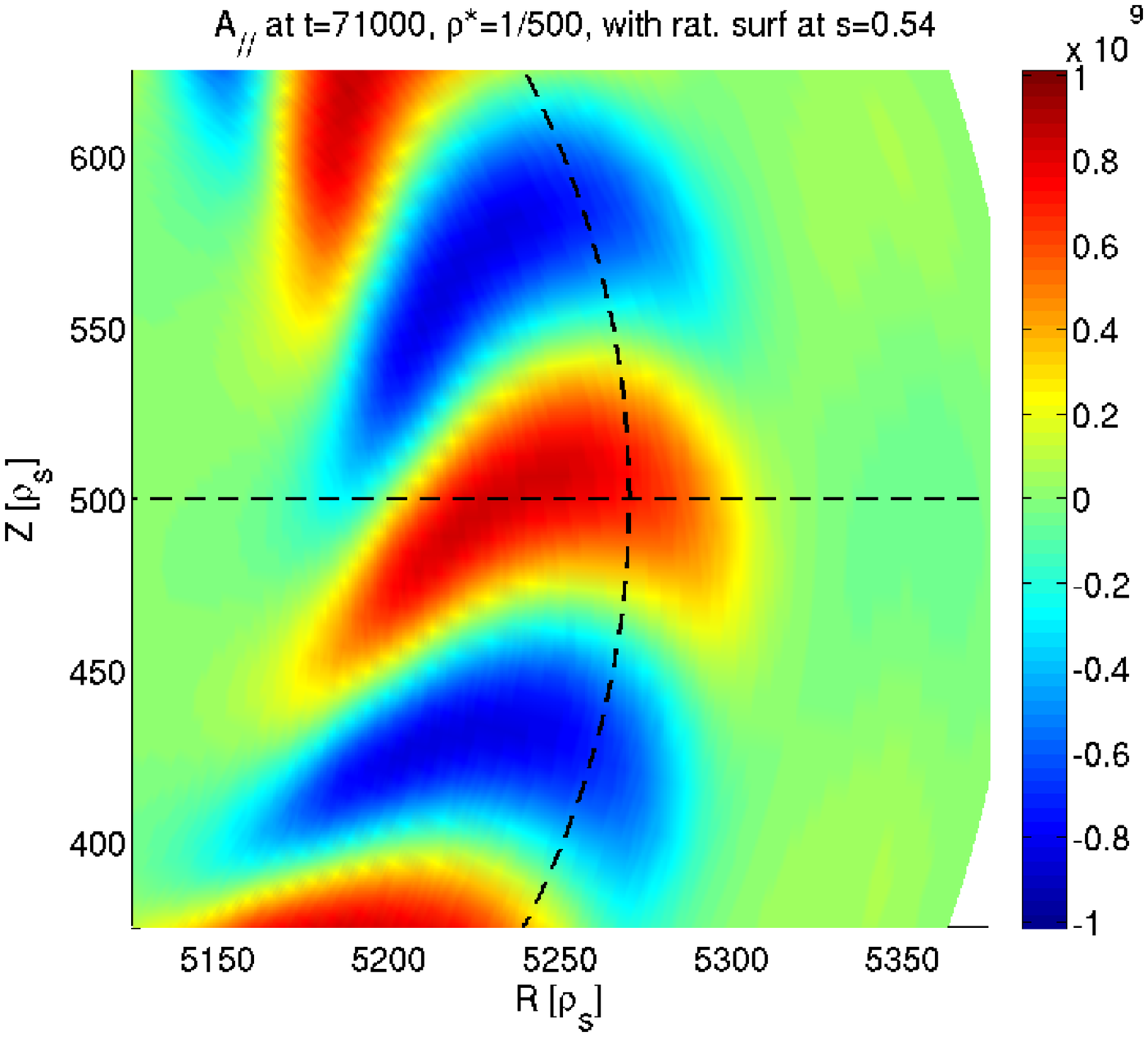}
\includegraphics[width=0.42\textwidth]{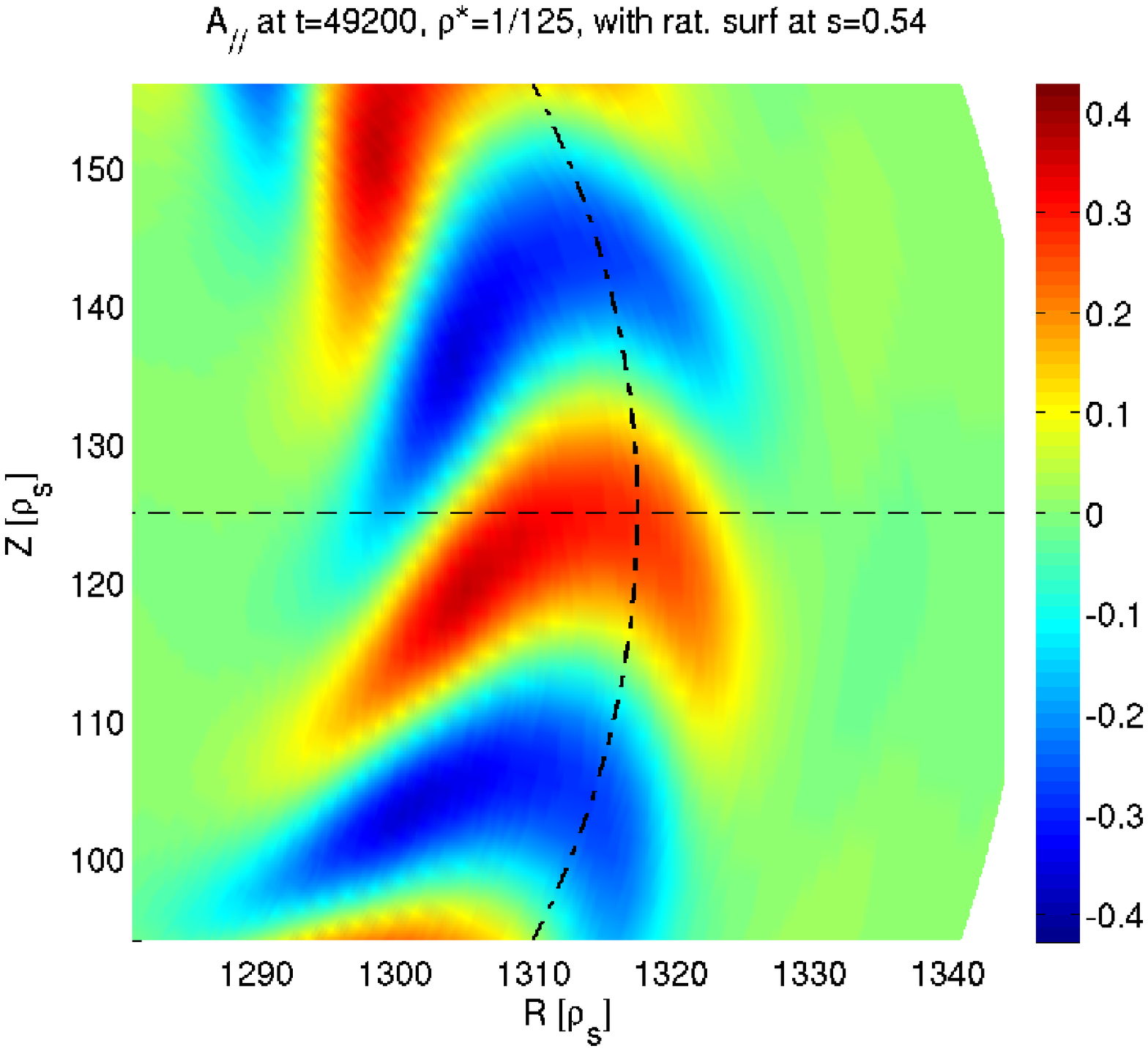}
\vskip -1.4em
\caption{Poloidal structure of $A_\|$ for the m=11 mode, for the reversed-shear q profile, and $n_{EP}/n_e =0.03$, $T_{EP}$=5510 keV. On the left, for $T_i$= 3.44 keV ($\rho^*=1/500$) and on the right for $T_i$=55.1 keV ($\rho^*=1/125$).}\label{fig:EPMm11-structure-revshearq}
\vskip -1em
\end{center}
\end{figure}

\begin{figure}[t!]
\begin{center}
\includegraphics[width=0.42\textwidth]{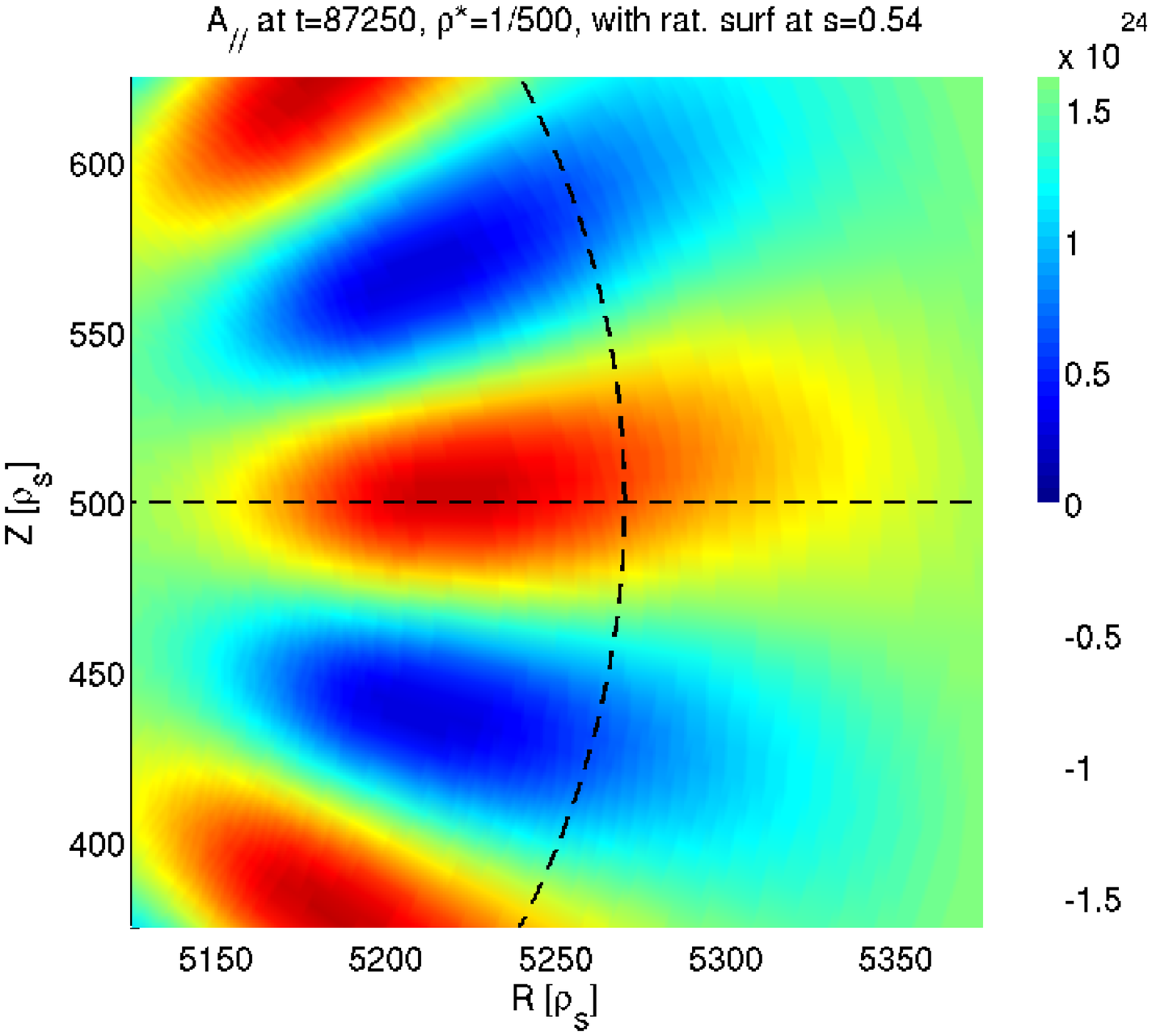}
\includegraphics[width=0.42\textwidth]{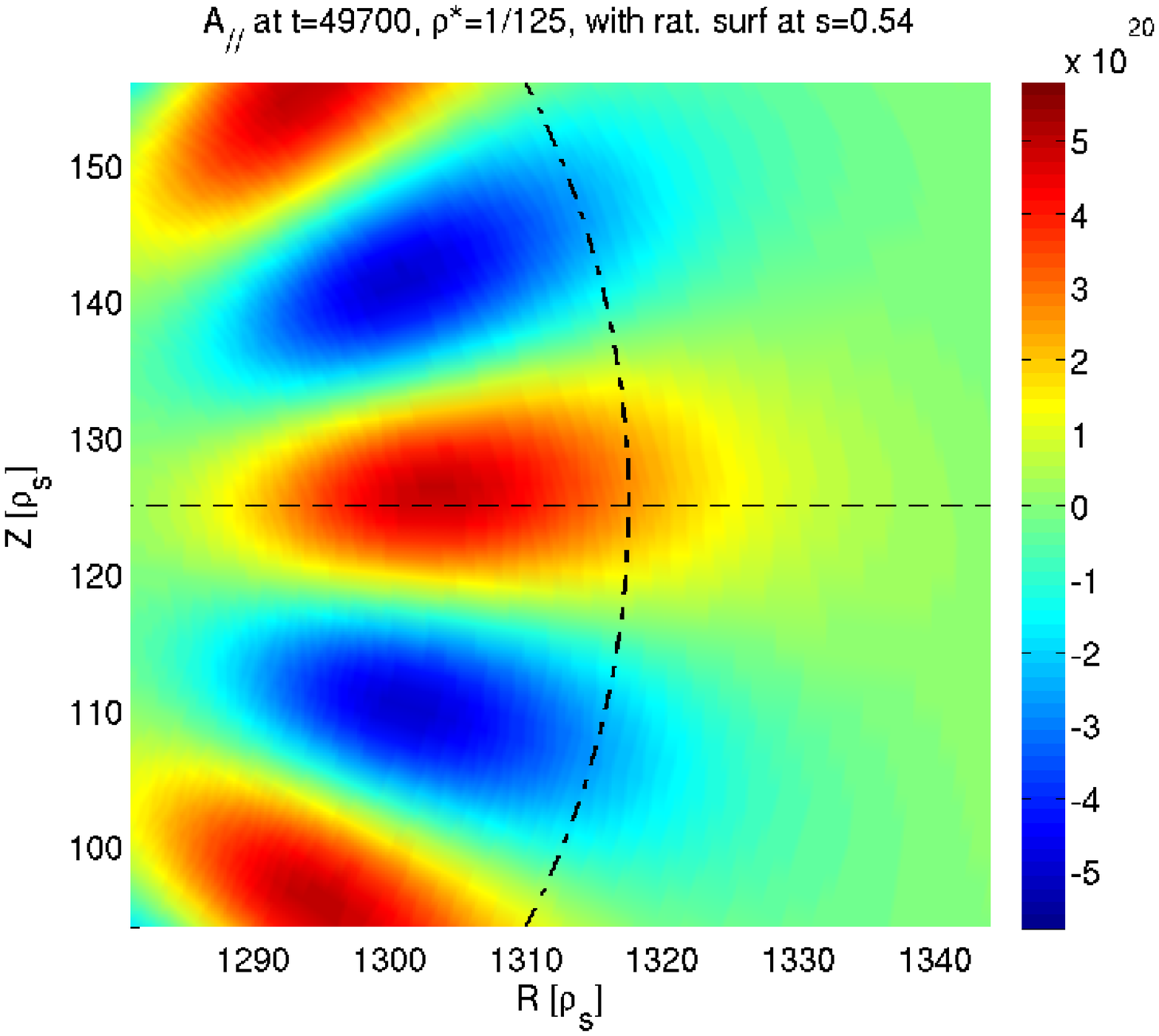}
\vskip -1.4em
\caption{Poloidal structure of $A_\|$ for the m=11 mode, for a case with flat q profile (q=1.78) (with $n_{EP}/n_e =0.03$, $T_{EP}$=5510 keV). On the left, for $T_i$= 3.44 keV ($\rho^*=1/500$) and on the right for $T_i$=55.1 keV ($\rho^*=1/125$).}\label{fig:EPMm11-structure-flatq}
\vskip -1em
\end{center}
\end{figure}


\begin{figure}[t!]
\begin{center}\label{fig:EPM-structure}
\includegraphics[width=0.42\textwidth]{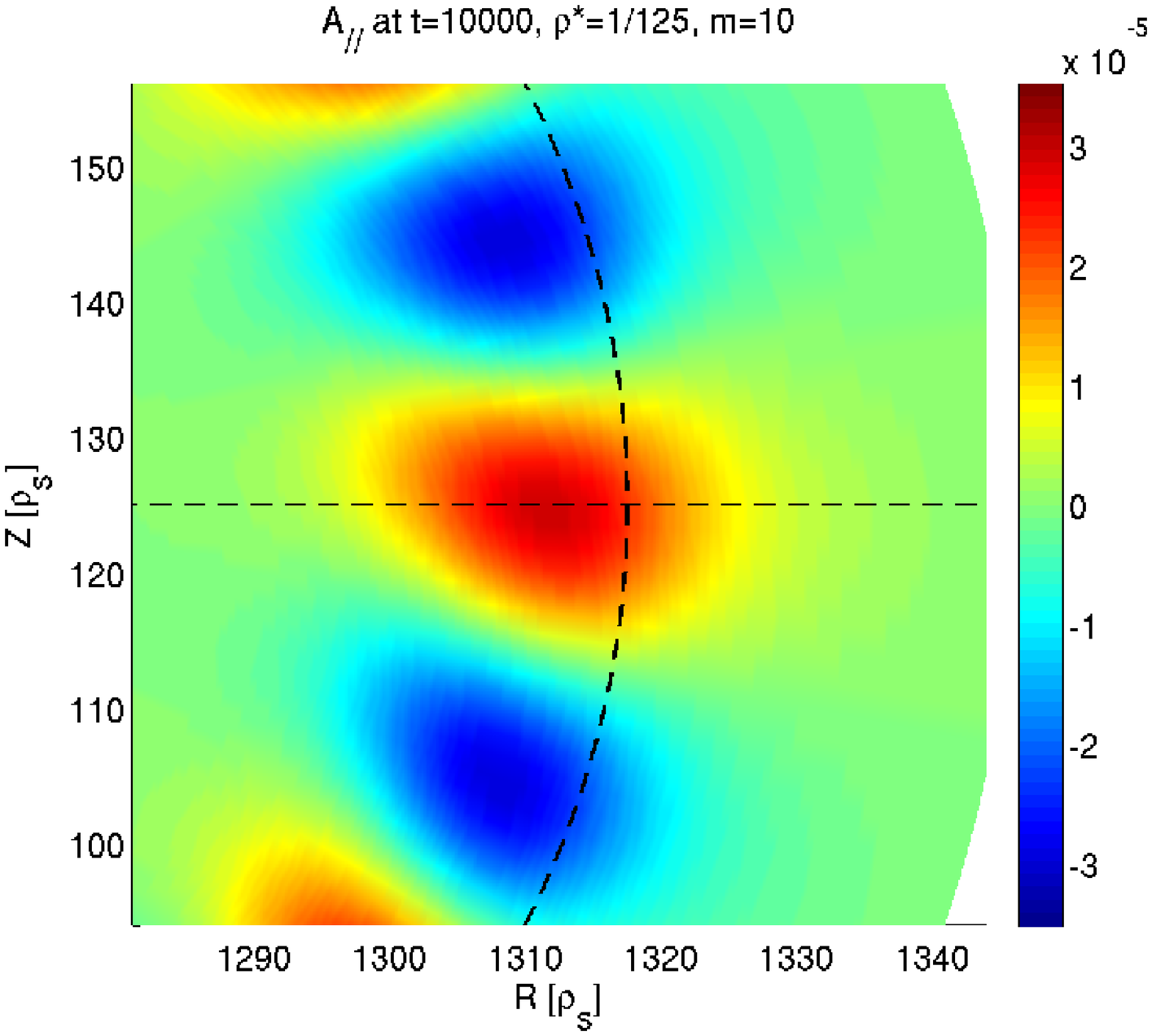}
\includegraphics[width=0.42\textwidth]{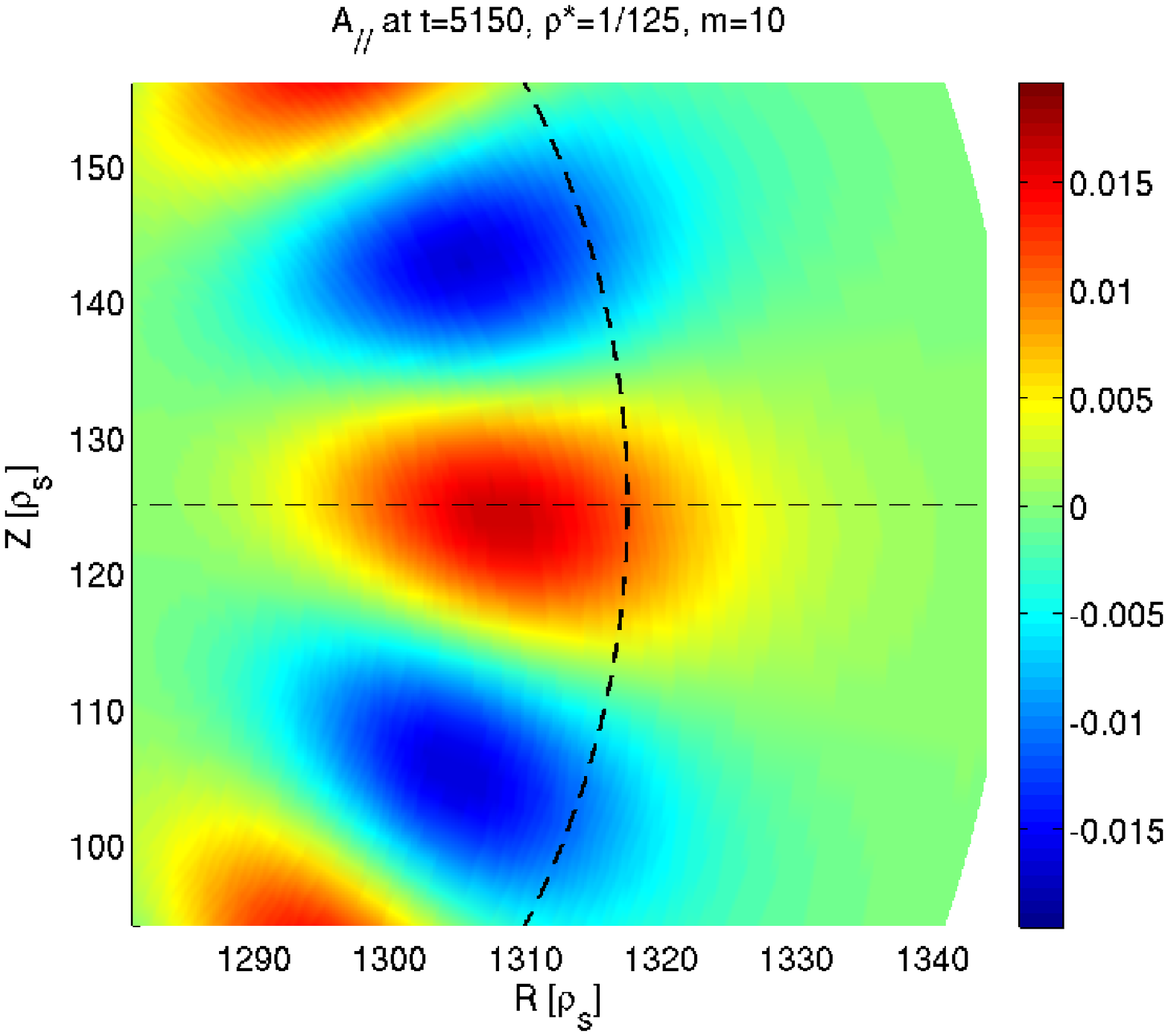}
\vskip -1.4em
\caption{Poloidal structure of $A_\|$ for the m=10 mode, for the case with reversed-shear q profile shown in Fig.~\ref{fig:EPM-q}, with $T_i$=55.1 keV, $T_{EP}$=5510 keV, and respectively $n_{EP}/n_e =0.01$ (left) and  $n_{EP}/n_e =0.03$ (right). Like for the m=11 mode, EP concentration seems to influence only the radial mode extension.}\label{fig:EPMm10-structure-revshearq}
\vskip -1em
\end{center}
\end{figure}

The mode is observed to rotate in the poloidal angle with amplitude increasing exponentially in time and well defined growth rate (described in the previous section). For the mode with m=11, a strong phase shift is observed at lower and bigger radii w.r.t. the signal measured at the CAP flux surface, which is visualized as a tilt of the poloidal structure at both sides, creating a ``boomerang'' shape (see also e.g. Ref.~\cite{Bass10,Deng10,Tobias11,Ma15}).
The tilt angle at both inner and outer radii is not observed to depend on the EP concentration, which seems to affect only the radial extension of the mode (see Fig.~\ref{fig:EPMm11-structure-vs_nEP}).
We also compare the structure of the m=11 mode for two cases with different bulk-ion temperature. Even in this case, the geometry is found not to change in the two cases (see Fig.~\ref{fig:EPMm11-structure-revshearq}).
Similarly, no evident difference in the poloidal structure is found by switching on/off the bulk-ion FLR (i.e. by actually calculating the correct value of the gyroaverage operators for the bulk-ions, for their finite value of $k_\perp \rho_i$).

The poloidal mode structure for the mode with m=11 is found to be modified, on the contrary, by the continuous spectrum topology. This is investigated by repeating the same case as in Fig.~\ref{fig:EPMm11-structure-revshearq} (for both bulk-ion temperatures) but for an equilibrium where the poloidal magnetic field is chosen such as to have a flat-q profile (i.e. zero shear), by keeping all other parameters unchanged. The result is shown in Fig.~\ref{fig:EPMm11-structure-flatq}, for a case with $T_i$= 3.44 keV and a case with $T_i$=55.1 keV.
In this case with flat-q profile, no tilt is found in the poloidal mode structure. The difference with respect of the reversed-shear case, is that in the flat-q case the bulk plasma nonuniformity (reflected in the continuum slope away from the CAP) is pushed down to zero, and the mode is not affected by continuum damping at any radial position.
Consistently, these m=11 modes are found to have a higher growth rate (e.g. $\gamma \simeq 2 \cdot 10^{-2} v_A/R$ for $n_{EP}/n_e=0.03$) w.r.t. the ones observed in the reversed-shear equilibrium. This is due to the absence of continuum damping, for this case where the mode is not touching a sloped continuum, at any radial location.

In order to complete the study of the dependence of the structure on the position w.r.t the continuum, we investigate also the poloidal structure of the m=10 mode, for $n_{EP}/n_e =0.01$ and  $n_{EP}/n_e =0.03$ (see Fig.~\ref{fig:EPMm10-structure-revshearq}). 
No strong tilt is found in any case (the mode is in the gap), although a slight upward tilt can be observed for the case with lower EP concentration (the mode is closer to the CAP).
In summary, we conclude that the tilt of the EP driven modes in this regime strongly depends on the slope of the continuum, for those modes which strongly interact with the continuum, being absent at the location where the continuum slope goes to zero or where there is no interaction with the continuum (i.e. for those cases where the continuum damping does not affect the modes).

The dependence of the poloidal mode structure of Alfv\'en instabilities on the features of the EP population (radial location of the density gradient, value of the density gradient, temperature, etc.) and the comparison with analytical theory (see e.g. Ref.~\cite{Ma15}), is outside the scope of this paper, and will be investigated in details in a dedicated paper.



\section{Summary and conclusions.}
\label{sec:conclusions}

The importance of understanding the shear-Alfv\'en wave (SAW) dynamics is mainly linked to their role in the redistribution of the energetic particle (EP) population. This is crucial for the achievement of a good theoretical model of the plasma stability and heating. The linear and nonlinear interaction of SAW instabilities with EP and with other modes (e.g. with zonal flows) make their investigation not trivial, for many space and time scales become involved and the nonlinearities and the driving and dissipation mechanisms must be treated as rigorously as possible.
To face such a complicated system, a robust theoretical tool is required with a set of model nonlinear equations constructed in such a way to conserve the basic symmetries of the system (energy and momentum).

We adopt here the code ORB5 within the NEMORB project, which has been previously used for turbulence simulations, and for the study of electrostatic global instabilities driven by EP. NEMORB's model equations are derived in a gyrokinetic Lagrangian formulation, where the discretization is performed at the Lagrangian level, so that the Vlasov-Maxwell governing equations satisfy the same symmetry properties of the starting discretized Lagrangian.

In this paper, we have presented the results of the first investigation performed with NEMORB on the linear collisionless dynamics of SAW instabilities. Firstly, the model equations have been presented and solved analytically for radially localized modes, in the incompressible ideal MHD limit. This gives the frequency of the SAW continuous spectrum, which is the local oscillation frequency of SAW in a tokamak. The continuous spectrum also provides the position of energy absorption of global SAW modes via continuum damping, and its topology is therefore important to know when studying the existence of global SAW instabilities.

As a first test of NEMORB on local SAW dynamics, we have performed numerical simulations in simplified geometry with negligible inverse aspect ratio, in order to recover the cylindrical limit. The frequency of the SAW oscillation has been measured and compared with the theoretical prediction for axisymmetric and non-axisymmetric perturbations. No EP population has been loaded at this stage.

The dynamics of global SAW instabilities has also been investigated, e.g. for Toroidicity induced Alfv\'en Eigenmodes driven by an EP population. The frequency of TAE with respect to the theoretical continuous spectrum has been investigated and the growth rate dependence with respect to the EP temperature has been studied. Results have also been compared with those of the hybrid gyrokinetic-MHD code HMGC, giving a good match.

Finally, we have investigated the dynamics of global SAW instabilities centered in a region with no magnetic shear. For this tokamak equilibrium configuration, the SAW frequency has been verified to tend to the prediction for the continuum, in the limit of vanishing EP concentration, and to decrease for increasing EP concentrations.
The dependence of the spatial structure in the poloidal plane on the equilibrium parameters has been investigated in details.
A phase shift in the poloidal angle $\theta$ at different radii has been observed, giving a characteristic ``boomerang'' shape. The radial size of the mode has been observed to depend on the EP concentration.
The shape has been observed to be directly linked to the position of the mode frequency with respect to the continuous spectrum: modes with frequency in the continuum gap have no phase shift in $\theta$, i.e. no boomerang shape, whereas modes entering the continuum for increasing EP concentration have a well defined   boomerang shape.
The investigation of the dependence of the spatial structure on the EP distribution function is outside the scope of this paper, and will be investigated with NEMORB in a dedicated paper and the results will be compared with analytical theory.

%

\section*{Acknowledgments.}

This work has been carried out within the framework of the EUROfusion Consortium and has
received funding from the Euratom research and training program 2014-2018 under grant
agreement No 633053. The views and opinions expressed herein do not necessarily reflect
those of the European Commission. Simulations were performed on the IFERC-CSC Helios
supercomputer within the framework of the ORBFAST and VERIGYRO projects.
This work has been done in the framework of the nonlinear energetic particle dynamics (NLED) European Enabling Research Project (EUROFUSION WP15-ER-01/ENEA-03), and  European Enabling Research Project on verification and development of new algorithms for gyrokinetic codes (EUROFUSION WP15-ER-01/IPP-01). Interesting discussions with L. Chen, M. Cole, Z. Qiu, C. Di Troia, G. Vlad, and A. Zocco are gratefully acknowledged. Part of this work was done when one of the authors, A. Biancalani, was visiting IPP at Greifswald, IFTS at Hangzhou, and ENEA at Frascati, whose teams are gratefully acknowledged for the hospitality. This paper has been written in Rochefort (France).

\end{document}